\documentclass[journal]{IEEEtran}
\IEEEoverridecommandlockouts
\usepackage{graphicx}
\usepackage{amssymb}
\usepackage{multirow}
\usepackage{indentfirst}
\usepackage{inputenc}
\usepackage{array}
\usepackage{stfloats}
\usepackage{multicol}
\usepackage[switch,mathlines]{lineno}
\usepackage{soul}
\usepackage{graphicx}
\usepackage{cite}
\usepackage[cmex10]{amsmath}
\newcounter{mytempeqncnt}
\usepackage{algorithmic}
\usepackage{array}
\usepackage{CJK}
\usepackage[cmex10]{amsmath}
\usepackage{algorithmic}
\usepackage{array}
\usepackage{fixltx2e}
\usepackage{bm}
\usepackage{amssymb}
\usepackage{color}
\usepackage{xcolor}

\usepackage{etoolbox}
\usepackage{nomencl}
\usepackage{setspace}
\usepackage{mathtools}
\usepackage{cuted} 
\usepackage[ruled, linesnumbered]{algorithm2e}
\ifCLASSOPTIONcompsoc
 \usepackage[caption=false,font=normalsize,labelfont=sf,textfont=sf]{subfig}
\else
 \usepackage[caption=false,font=footnotesize]{subfig}
\fi
\usepackage{url}
\usepackage{makecell}
\usepackage{colortbl}
\usepackage{color}
\usepackage[table,xcdraw]{xcolor}
\def\BibTeX{{\rm B\kern-.05em{\sc i\kern-.025em b}\kern-.08em
    T\kern-.1667em\lower.7ex\hbox{E}\kern-.125emX}}
\usepackage{balance}
\usepackage{booktabs, tabularx}
\usepackage[referable]{threeparttablex}
\usepackage{stfloats}
\usepackage{multirow}
\usepackage{float}

\usepackage[
top=1.7cm,
bottom=1.7cm,
left=1.5cm,
right=1.5cm
]{geometry}

\newcommand{\rv}[1]{\textcolor{black}{#1}}
\newcommand{\rvb}[1]{\textcolor{black}{#1}}

\begin{document}

\title{Local Energy Function-Based Power System Transient Stability Analysis with Grid Forming and Grid Following Inverters}

\author{{Yingyi~Tang,~\IEEEmembership{Student Member,~IEEE}, Junbo~Zhao,~\IEEEmembership{Senior Member,~IEEE}, Baosen Zhang,~\IEEEmembership{Member,~IEEE}
}
  \thanks{This work is supported in part by the U.S. Department of Energy Office of Electricity Advanced Grid Modeling program and Solar Energy Technologies Office. Y. Tang and J. Zhao are with Thayer School of Engineering, Dartmouth College, Hanover, NH 03755. B. Zhang is with the University of Washington, Seattle, WA 98195.}}


\maketitle
\begin{abstract}
The widespread deployment of inverter-based resources (IBRs) is reshaping power system dynamics by their control strategies. Grid-following (GFL) and grid-forming (GFM) inverters are expected to co-exist from now on. However, ensuring the transient stability of such systems remains a critical and underexplored challenge. In this paper, a comprehensive transient stability analysis of the mixed-GFM-GFL inverter system is conducted using a newly derived local energy function. The transient stability analysis model of the system is first derived to reveal the dynamic interactions between GFL and GFM inverters during transient responses. A local energy function is then proposed for analytical transient stability assessment. \rv{Instead of ignoring or making harsh assumptions on inverters' coupled damping effect, the proposed local energy function carefully addresses its impact on system transient stability by locally bounding the region where the system's energy dissipation property holds. Such a property fully considers the combined damping effect of GFL and GFM inverters.} The derived critical energy and stability index provide a fast and quantitative transient stability assessment. Furthermore, the adverse impact of insufficient damping of GFM inverters on system transient stability is thoroughly investigated. The effectiveness of the proposed method is validated on the modified IEEE 14-bus system with mixed GFL and GFM inverters, \rv{and the numerical simulation results further demonstrate the small conservativeness of the proposed method.}

\end{abstract}
\vspace{-0.3cm}
\section{Introduction}
\vspace{-0.1cm}

\IEEEPARstart{W}{ith} the increasing integration of inverter-based resources (IBRs), the dynamic behavior of power systems is undergoing a fundamental transformation \cite{Matevosyan_Julia}. Unlike traditional synchronous generators (SGs), the dynamics of IBRs are primarily governed by their control strategies rather than electromechanical characteristics \cite{Wang_Xiongfei}. Among the various control schemes, grid-following control (GFL) and grid-forming control (GFM) have emerged as the two most widely adopted approaches \cite{Li_Yitong}. Currently, most inverters deployed in power systems operate under GFL control \cite{Gu_Yunjie}, while recent research has increasingly focused on GFM inverters due to their ability to support the system stability. 

GFL inverters use a phase-locked loop (PLL) to synchronize with the grid \cite{Huang_Liang}. This makes them particularly effective in scenarios, where grid stability is well-maintained and the system operates under strong grid conditions \cite{Geng_Hua}. In contrast, GFM inverters do not depend on an external grid for synchronization as they can establish and regulate system voltage and frequency by emulating the behavior of traditional SGs \cite{Kerdphol, Liu_Hao, Yazdani}. This capability makes GFM inverters particularly advantageous in inverter-dominated systems \cite{Fu_Xikun}. 

Despite the advantages of GFM inverters, retrofitting existing inverters to GFM inverters is currently neither economically feasible nor practically viable \cite{Zhang_Haobo,Lasseter_Robert}. Consequently, a mixed approach incorporating a combination of GFL and GFM inverters, is likely to emerge as a predominant solution. However, the mixed-GFM-GFL-inverter systems exhibit highly nonlinear and intricate dynamics \cite{Fu_Xikun2}. This interaction between different types of inverters has a critical impact on overall system stability. 

Existing stability studies of the mixed-GFM-GFL inverter system can be mainly divided into small-signal stability and transient stability. Great efforts have been made to analyze the small signal stability of the mixed GFM-GFL inverter system \cite{Xin_Huanhai, Shi_Qianhong, Ding_Lizhi, Xin_Huanhai2, Singh, Gao_Yu, Li_Xiayan}, offering critical insights into the spectra of the eigenvalue of the system, the oscillation modes and the parameter sensitivity associated with the dynamics of the inverter. While the small-signal analysis provides important insights into system characteristics and parameter tuning, it inherently relies on linearized models around operating points. As a result, it cannot fully capture the system’s behavior under large disturbances, such as faults, sudden disconnections, or sharp load changes. Moreover, the effectiveness of GFM control in enhancing small-signal stability raises a question: is the mere addition of GFM inverters sufficient to ensure system transient stability, or are further conditions required to secure stable system performance following large disturbances? Inverter-dominated systems, particularly those with strong GFM-GFL interactions and low inertia, can exhibit complex nonlinear behaviors that are invisible to small-signal methods. This motivates a deeper investigation into transient stability, where the system's ability to maintain synchronization after large disturbances is further evaluated.

In recent studies, the transient stability of mixed GFM-GFL-inverter system has been analyzed using models of varying orders and different methods, such as phase portrait method, equal area criterion (EAC) method and Lyapunov's method. In \cite{Wang_Shunliang}, a third-order model and a 2D phase portrait method based on dimensionality reduction were proposed, indicating that reducing the inverter active power setpoint improves stability. However, the method is limited to two-inverter systems and lacks scalability. \cite{Lu_Dian} introduced a fourth-order synchronization model considering controller interactions and applied phase portrait method to compare different controller interactions' impact on synchronization process. Yet, the phase portrait remains a numerical tool, limiting parameter-level insights. 

Considering the impact of the injected current from GFL inverter in mixed multi-inverter system, \cite{Li_Mingfei} suggested that increased GFL terminal current supports GFM stability, whereas changes in GFM voltage shift GFL equilibrium point. These conclusions are drawn primarily from an intuitive interpretation of the system model's mathematical expressions, without a rigorous theoretical foundation. In contrast, \cite{Zhang_Xiaonan} used the EAC method to argue that GFL current injection can suppress GFM output under large disturbances, degrading stability, especially under current saturation. The EAC method was also adopted in \cite{Li_Nan} \rv{and \cite{Me_Si_Phu} to analyze power angle interactions, where corresponding control strategies were proposed to stabilize the system in each study. Similarly, \cite{He_Yifan} applied the EAC method to investigate the GFM inverter's impact on the GFL inverter's transient stability by comparing acceleration and deceleration areas. Furthermore, \cite{Li_Zhijie} investigated the system acceleration and deceleration areas under different control coefficients and proposed an optimal parameter selection method.}
However, because the EAC method assumes positive damping, the impact of the negative damping brought by PLL control on system transient stability cannot be fully investigated.

Regarding the Lyapunov function method, whether in quadratic form or energy-function form, a central challenge lies in how to construct an appropriate Lyapunov function for this nonlinear, high-order and coupled system. In \cite{Wang_Yang}, a Lyapunov function was constructed in standard quadratic form using a Takagi–Sugeno fuzzy model. However, without analytical expressions, the parameter impacts on the estimated stability region were assessed only iteratively.

For energy function based transient stability analysis, the classical direct method defines an energy function as a scalar Lyapunov-like function $V(x)$, typically constructed via the first integrals of system motion, whose time derivative satisfies $\dot{V}(x)\le 0$ along any system trajectories \cite{Sauer_Peter_W}. \rvb{Although both the energy function method and the Lyapunov function method certify stability through a scalar function that decays along trajectories, they differ in important respects. An energy function admits a direct physical interpretation as the sum of kinetic and potential energy of the system, derived structurally from the equations of motion via first integrals, whereas a Lyapunov function is a mathematical construct chosen primarily to satisfy the Lyapunov condition, often without an underlying physical meaning~\cite{Athay1979,Chiang}. Moreover, the energy function method provides direct access to the relevant unstable equilibrium
points (UEPs) characterizations of the stability boundary~\cite{Chiang}, in which the critical energy is identified with the energy at a specific UEP. Lyapunov methods, by contrast, typically delineate the stability region as a constant level set $\{V \le c\}$ whose level value $c$ is determined by the largest sublevel set on which the LaSalle invariance condition holds~\cite{Khalil2002}, without reference to any physically meaningful escape mechanism.} A global energy function further requires that $\dot{V}(x)\le 0$ hold over the entire state space. When such a global energy function exists for the studied power system model, the stability boundary can be characterized by the union of the stable manifolds of the UEPs \cite{Rimorov_Dmitry}. \cite{Huang_Meng} proposed an energy function for GFL inverter and integrated the GFM inverter's influence into the parameter-dependent domain of attraction (PDA). However, the GFM inverter's impact is not explicitly quantified and requires repeated plotting of the PDAs under different GFM operating conditions, which can be computationally intensive and lacks direct analytical insight.
In \cite{Tian_Zhen}, a Hamiltonian energy function was constructed for a two-inverter islanded microgrid. By setting the GFM inverter as the reference bus, this approach resulted in the GFM inverter's dynamics not being fully incorporated into the stability region estimation. Similarly, \cite{Qu_Qiannan} developed an energy function for a mixed GFM-GFL-inverter system based on the definition of damping energies. Although this work further examined the damping effect of the GFM inverter, the damping contributions of the GFM and GFL inverters were analyzed separately, and the derivation of the stability criterion requires both damping terms to be negative, leading to the conservativeness of the proposed method.

The above studies have contributed to understanding the dynamics of mixed GFM-GFL-inverter systems. However, the analytical transient stability analysis of such systems remains insufficiently addressed due to the stringent requirement of a global energy function that satisfies $\dot{V}(x)\le 0$ pointwise over the entire state space. \rvb{To address this challenge, this paper adopts a local energy function framework, which relaxes this classical constraint by only requiring $\dot{V}(x)\le 0$ to hold outside a small bounded region surrounding the post-fault stable equilibrium point (SEP).  Mathematically, this localized energy dissipation property establishes uniform ultimate boundedness of post-fault trajectories within a residual neighborhood of the SEP, which is denoted as Lagrange stability \cite{El_abiad}. Crucially, the radius of this residual neighborhood is not a fixed quantity but contracts to zero as the trajectory approaches the SEP. Consequently, the ultimate-boundedness conclusion strengthens into asymptotic convergence to the SEP, and the proposed framework delivers transient stability in the standard power-system engineering sense.} The main contributions of this work are summarized as follows:

1) The mathematical model of the mixed-GFM-GFL-inverter system is established, where the inverters are not required to share the same point of connection. The derived inter-inverter coupling terms reveal how the PLL controllers of GFL inverters and VSG controllers of GFM inverters interact with each other.

2) A local energy function jointly leveraging the damping effect from GFL and GFM inverters is derived for the mixed-GFM-GFL-inverter system, allowing for a more comprehensive transient stability analysis. The system critical energy is further derived for fast transient stability assessment, and a corresponding stability index is proposed to provide a quantitative stability measure after large disturbances.

3) The adverse impact of poorly damped GFM inverters on system transient stability is thoroughly investigated. It is demonstrated that inadequately damped GFM inverters can degrade the damping performance of GFL inverters through inter-inverter coupling, thereby negatively affecting overall system stability.

\vspace{-0.3cm}
\section{Mathematical Model of Mixed-GFM-GFL-Inverter System}
\subsection{AC Network Modeling with GFL and GFM Inverters}
 \begin{figure}[!t]
\centering
\vspace{-0.1cm}
\includegraphics[width=3.3in]{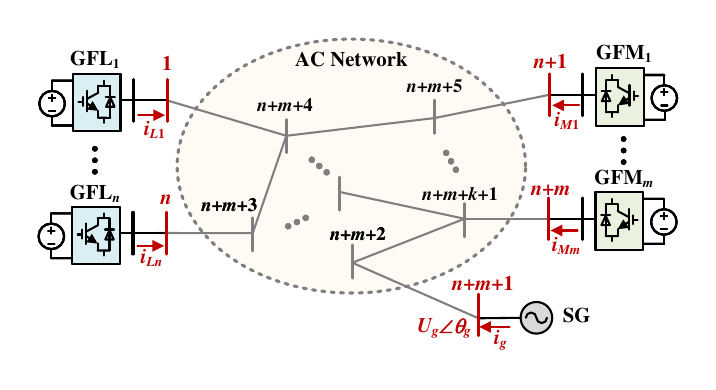}
\vspace{-0.3cm}
\caption{Transmission system incorporated with mixed GFM and GFL inverters.}
\vspace{-0.3cm}
\label{Fig1}
\end{figure}
As shown in Fig. \ref{Fig1}, $n$ GFL inverters (Bus 1 $\sim$ Bus $n$) and $m$ GFM inverters (Bus $n+1$ $\sim$ Bus $n+m$) are integrated into the power network. The synchronous generator at Bus $n+m+1$ has comparably large inertia and is thereby denoted as an infinite voltage source $U_ge^{j\theta_g}$ \cite{Li_Yujun}. Consider that there are $k$ interior non-sourced buses (Bus $n+m+2$ $\sim$ Bus $n+m+k$) in the power network. Regarding the boundaries buses (subscript $\bm{\alpha}$) and the interior buses (subscript $\bm{\beta}$), the network admittance matrix $\mathbf{Y}$ can be divided into submatrices as follows:
\begin{equation}
    \left[ \begin{matrix}
   \mathbf{i}_{\bm{\alpha }}  \\
   \mathbf{i}_{\bm{\beta }}  \\
\end{matrix} \right]=\left[ \begin{matrix}
   \mathbf{Y}_{\bm{\alpha}} & \mathbf{Y}_{\bm{\alpha \beta}} \\
   {\mathbf{Y}_{\bm{\beta \alpha}}} & \mathbf{Y}_{\bm{\beta}}  \\
\end{matrix} \right]    \left[ \begin{matrix}
   \mathbf{u}_{\bm{\alpha }}  \\
   \mathbf{u}_{\bm{\beta }}  \\
\end{matrix} \right]
\label{eq1}
\end{equation}
where $\mathbf{i}_{\bm{\alpha}},\ \mathbf{u}_{\bm{\alpha}} \in {{\mathbb{R}}^{(n+m+1)\times 1}}$ are the current and voltage of the buses integrated with inverters and SG. $\mathbf{i}_{\bm{\beta}},\ \mathbf{u}_{\bm{\beta}} \in {{\mathbb{R}}^{k\times 1}}$ are the injection current and voltage of the non-sourced buses. $\mathbf{Y}_{\bm{\alpha}}\in {{\mathbb{R}}^{(n+m+1)\times (n+m+1)}}$,  $\mathbf{Y}_{\bm{\alpha \beta}}\in {{\mathbb{R}}^{(n+m+1)\times k}}$, $\mathbf{Y}_{\bm{\beta \alpha}}\in {{\mathbb{R}}^{k\times (n+m+1)}}$ and $\mathbf{Y}_{\bm{\beta}}\in {{\mathbb{R}}^{k\times k}}$. The constant impedance loads are incorporated into the admittance matrix. 

Applying Kron reduction with $\mathbf{i}_{\bm{\beta }} =\mathbf{0}$, $\mathbf{i}_{\bm{\alpha}}$ can be derived based on (\ref{eq1}) as $\mathbf{i}_{\bm{\alpha }}=\left( \mathbf{Y}_{\bm{\alpha}}-\mathbf{Y}_{\bm{\alpha \beta}}\mathbf{Y}_{\bm{\beta}}^{-1}\mathbf{Y}_{\bm{\beta \alpha}} \right)\mathbf{u}_{\bm{\alpha }}$. Thus we define:
\begin{equation}
    \mathbf{Y}_{\mathbf{new}}=\mathbf{Y}_{\bm{\alpha}}-\mathbf{Y}_{\bm{\alpha \beta}}\mathbf{Y}_{\bm{\beta}}^{-1}\mathbf{Y}_{\bm{\beta \alpha}}.
\label{eq:Ynew}
\end{equation}
\rv{where the load characteristics are primarily conserved within the self-admittance terms of $\mathbf{Y_{new}}$.} 

\rv{It is worth noting that after Kron reduction, the algebraic constraints of the network are eliminated to expose the dominant synchronization dynamics among inverter nodes, which is a common and well-established practice in both classical transient stability analysis and IBR modeling. Importantly, the reduced network preserves the steady-state operating point and the fundamental power–angle relationships of the retained buses, which are the primary variables governing transient synchronization.}

The new admittance matrix after Kron reduction $\mathbf{Y}_{\mathbf{new}}$ can be further written in submatrices regarding the GFL inverter buses, the GFM inverter buses and the SG bus as:
\begin{equation}
       \left[ \begin{matrix}
   \mathbf{i}_{\mathbf{L}}  \\
   \mathbf{i}_{\mathbf{M}}  \\
   i_g  \\
\end{matrix} \right]=\mathbf{Y}_{\mathbf{new}}\left[ \begin{matrix}
   \mathbf{u}_{\mathbf{L}}  \\
   \mathbf{u}_{\mathbf{M}}  \\
   u_g  \\
\end{matrix} \right]=\left[ \begin{matrix}
   \mathbf{Y}_{\mathbf{L}} &  \mathbf{Y}_{\mathbf{LM}} & \mathbf{Y}_{\mathbf{Lg}} \\
   \mathbf{Y}_{\mathbf{ML}} &  \mathbf{Y}_{\mathbf{M}} & \mathbf{Y}_{\mathbf{Mg}} \\
   \mathbf{Y}_{\mathbf{gL}} &
   \mathbf{Y}_{\mathbf{gM}} &
   \mathbf{Y}_{\mathbf{g}}  \\
\end{matrix} \right]\left[ \begin{matrix}
   \mathbf{u}_{\mathbf{L}}  \\
   \mathbf{u}_{\mathbf{M}}  \\
   u_g  \\
\end{matrix} \right]
\label{eq:Ynewsb}
\end{equation}
where $[\mathbf{i}_{\mathbf{L}},\ \mathbf{i}_{\mathbf{M}},\ i_g]^\top=\mathbf{i}_{\bm{\alpha}}$ and $[\mathbf{u}_{\mathbf{L}},\ \mathbf{u}_{\mathbf{M}},\ u_g]^\top=\mathbf{u}_{\bm{\alpha}}$. $\mathbf{i}_{\mathbf{L}}, \ \mathbf{u}_{\mathbf{L}} \in {{\mathbb{R}}^{n\times 1}}$ are the current and voltage of GFL inverters,  $\mathbf{i}_{\mathbf{M}}, \ \mathbf{u}_{\mathbf{M}} \in {{\mathbb{R}}^{m\times 1}}$ are the current and voltage of GFM inverters, and $i_g,\ u_g$ are the current and voltage of the SG. 

\subsection{Dynamic Model of Coupled GFM and GFL Inverters for Transient Stability Analysis}
\begin{figure}[!t]
\vfil
\subfloat[GFL rotating reference frames]{\includegraphics[width=1.75in]{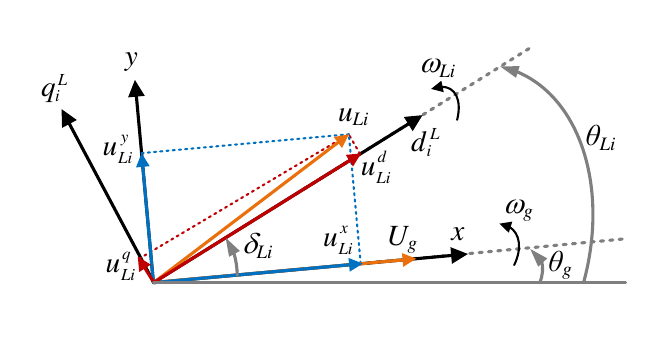}}%
\hfil
\hspace{-0.15in}
\subfloat[GFM rotating reference frames]{\includegraphics[width=1.75in]{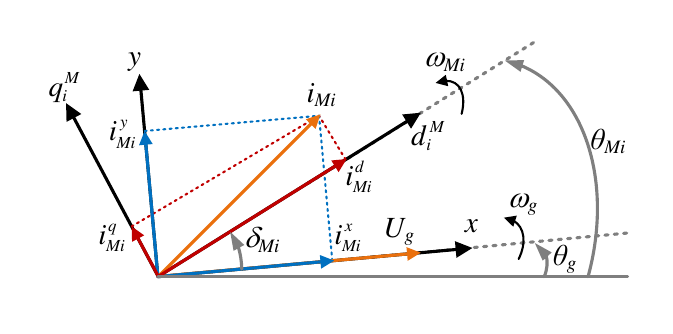}}%
\caption{$xy$ rotating reference frames and inverters $dq$ rotating reference frames. }
\vspace{-0.4cm}
\label{Fig2}
\end{figure}
As shown in Fig. \ref{Fig2}, the common $xy$ reference frame is rotating with the ideal angular frequency of $\omega_g$ (1 p.u.), the $i$-th GFL inverter's PLL $d_i^Lq_i^L$ reference frame is rotating with the angular frequency of $\omega_{Li}$ ($i=1,\dots, n$), and the $i$-th GFM inverter's $d_i^Mq_i^M$ reference frame is rotating with the angular frequency of $\omega_{Mi}$ ($i=1,\dots, m$). Then, $\theta_g$, $\theta_{Li}$ and $\theta_{Mi}$ are the angles of the SG, the $i$-th GFL inverter and the $i$-th GFM inverter respectively.

The $x$-axis of $xy$ rotating frame is aligned with the voltage of the infinite source $U_ge^{j\theta_g}$, thereby $u_g^x=U_g$, $u_g^y=0$, yielding: 
\begin{equation}
    u_g={{U}_{g}}{{e}^{j{{\theta }_{g}}}},\quad 
    i_g=(i_{g}^{x}+ji_{g}^{y}){{e}^{j{{\theta }_{g}}}}.
    \label{eq:ugig}
\end{equation}

Consequently, the point of common coupling (PCC) current and voltage of the $i$-th GFL inverter ($\mathbf{i}_{\mathbf{L}(i)}$, $\mathbf{u}_{\mathbf{L}(i)}$) and the $i$-th GFM inverter ($\mathbf{i}_{\mathbf{M}(i)}$, $\mathbf{u}_{\mathbf{M}(i)}$) are presented in $xy$ and $dq$ rotating frames as follows:
\begin{subequations}
\begin{equation}\left\{
\begin{aligned}
     &\mathbf{i}_{\mathbf{L}(i)}=
   (i_{Li}^{x}+ji_{Li}^{y}){{e}^{j{{\theta }_{g}}}}  
=(I_{Li}^{d}+jI_{Li}^{q}){{e}^{j{{\theta }_{Li}}}}\\
   & \mathbf{u}_{\mathbf{L}(i)}=
   (u_{Li}^{x}+ju_{Li}^{y}){{e}^{j{{\theta }_{g}}}}=
   (u_{Li}^{d}+ju_{Li}^{q}){{e}^{j{{\theta }_{Li}}}}  
\end{aligned} \right.
\label{eq:iLuL}
\end{equation}

\begin{equation}\left\{
\begin{aligned}
    &\mathbf{i}_{\mathbf{M}(i)}=
   (i_{Mi}^{x}+ji_{Mi}^{y}){{e}^{j{{\theta }_{g}}}} =
   (i_{Mi}^{d}+ji_{Mi}^{q}){{e}^{j{{\theta }_{Mi}}}} \\
       &\mathbf{u}_{\mathbf{M}(i)}=
   (u_{Mi}^{x}+ju_{Mi}^{y}){{e}^{j{{\theta }_{g}}}}  =
   (U_{Mi}^{d}+jU_{Mi}^{q}){{e}^{j{{\theta }_{Mi}}}} 
   \end{aligned} \right.
\label{eq:iMuM}
\end{equation}
\end{subequations}
where, after Kron reduction, all quantities are expressed in the common $xy$ reference frame aligned with angle $\theta_g$. As seen in  Fig. \ref{Fig2}, they can be converted into their local rotating $dq$ frame (with angle $\theta_i$) by the transformation $F_i^d+jF_i^q=(F_i^x+jF_i^y)e^{j(\theta_g-\theta_i)}$, where $F_i$ denotes the inverter's current or voltage. For notational simplicity, the superscripts $d$ and $q$ are not explicitly annotated to distinguish whether they correspond to the GFL or GFM inverter. Unless otherwise specified, the association of the superscripts $d$ and $q$ with either a GFL or GFM inverter is implicitly determined by the corresponding subscript of the variable.

\rv{Notably, for GFL inverters, the transient synchronization behavior is dominated by PLL dynamics as the bandwidth of the PLL is designed much lower than that of the inner current loop and the inner current control dynamics are typically neglected \cite{Zhou_Yunpeng,Hu_Qi_Fu_Lijun,Wu_Heng,Se_Kyo_Chung,Harnefors_Lennart2}.} Consequently, in (\ref{eq:iLuL}), the output currents of GFL inverters are considered to remain at their reference values: $i_{Li}^{d}=I_{Li}^{d}$, $i_{Li}^{q}=I_{Li}^{q}$ ($i=1,\dots, n$). 

\rv{Likewise, for GFM inverters, its transient synchronism is governed primarily by the power synchronization loop as its inner voltage/current control loops have bandwidths an order of magnitude higher than the power synchronization loop \cite{Luo_Ling, Shuai_Zhikang,Qu_Qiannan2,Rokrok_Ebrahim}.} Thus, in (\ref{eq:iMuM}), their PCC voltages are considered to remain at reference values: $u_{Mi}^{d}=U_{Mi}^{d}$, $u_{Mi}^{q}=U_{Mi}^{q}$ ($i=1,\dots, m$).

As shown in Fig. \ref{Fig2}, the power angles for GFL and GFM inverters are defined as their angle differences with $\theta_g$:
\begin{equation}
    {{\delta }_{Li}}={{\theta }_{Li}}-{{\theta }_{g}},\quad {{\delta }_{Mi}}={{\theta }_{Mi}}-{{\theta }_{g}}.
    \label{eq:power angle}
\end{equation}

\begin{figure}[!t]
    \centering

    \subfloat[PLL control frame for GFL inverters]{%
        \includegraphics[width=0.92\columnwidth]{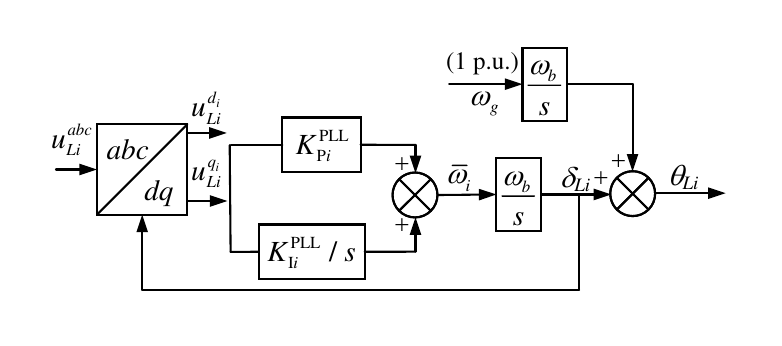}%
        \label{fig:control_gfl}%
    }

    \par\vspace{-2mm}

    \subfloat[VSG control frame for GFM inverters]{%
        \includegraphics[width=0.86\columnwidth]{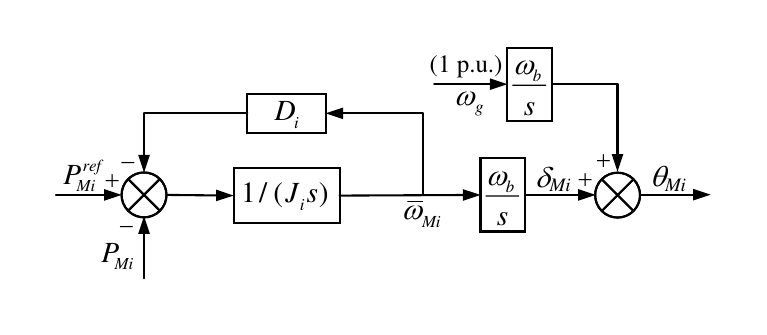}%
        \label{fig:control_gfm}%
    }

    \caption{Synchronous control frames of GFL and GFM inverters.}
    \label{Control}
\end{figure}

As shown in Fig. \ref{Control} (a), GFL inverter's PLL control operates by regulating inverter's $q$-axis PCC voltage to zero to synchronize with the grid: 
\begin{equation}
\left\{ \begin{aligned}
  & \dot{{\delta }}_{Li}=\omega_b \cdot {{{\bar{\omega }}}_{Li}} \\ 
 & \dot{{{\bar{\omega }}}}_{Li}=K_{\text{P}i}^\text{PLL}\dot{u}_{Li}^{q}+K_{\text{I}i}^\text{PLL}u_{Li}^{q} \\ 
\end{aligned} \right.
\label{eq:originPLL}
\end{equation}
where ${{{\bar{\omega }}}_{Li}}={{\omega }_{Li}}-{{\omega }_{g}}$. $K_{\text{P}i}^\text{PLL}$ and $K_{\text{I}i}^\text{PLL}$ are the proportional and integral gains of the $i$-th GFL inverter's PLL control, respectively.

GFM inverter's VSG control, as shown in Fig. \ref{Control} (b), emulating the synchronous machine swing equation by synchronize the inverter with the gird based on the difference of its output active power with the power reference:
\begin{equation}
    \left\{ \begin{aligned}
  & {{{\dot{\delta }}}_{Mi}}={{\omega }_{b}}\cdot {{{\bar{\omega }}}_{Mi}}\\ 
 & {{{\dot{\bar{\omega }}}}_{Mi}}=\frac{P_{Mi}^{\mathrm{ref}}-{{P}_{Mi}}}{{{J}_{i}}}-\frac{{{D_i}}}{{{J}_{i}}}{{{\bar{\omega }}}_{Mi}} \\ 
\end{aligned} \right.
\label{VSGcontrol}
\end{equation}
where ${{{\bar{\omega }}}_{Mi}}={{\omega }_{Mi}}-{{\omega }_{g}}$. $P_{Mi}^{\mathrm{ref}}$, $J_i$ and $D_i$ denote the reference active power, virtual inertia and damping coefficient of the $i$-th GFM inverter's VSG control. The active power of the $i$-th GFM inverter $P_{Mi}$ is calculated by $P_{Mi}=U_{Mi}^{d}i_{Mi}^{d}+U_{Mi}^{q}i_{Mi}^{q}=U_{Mi}^{d}i_{Mi}^{d}$, where the $q$-axis voltage reference of GFM inverters is normally set as zero.

\rv{For GFM inverters, droop control is also a commonly adopted synchronization method. In conventional droop control, a low-pass filter is usually added in the active power control loop to mitigate measurement noise. The dynamics of droop control with a low-pass filter can be described by
\begin{equation*}
 \begin{aligned}
   & \left\{ \begin{aligned}
  & {{{\dot{\delta }}}_{Mi}}={{\omega }_{b}}\cdot {{{\bar{\omega }}}_{Mi}}={{\omega }_{b}}({{\omega }_{Mi}}-{{\omega }_{g}}) \\ 
 & {{{\bar{\omega }}}_{Mi}}={{k}_{pi}}\cdot \frac{{{\omega }_{ci}}}{s+{{\omega }_{ci}}}\left( P_{Mi}^\text{ref}-{{P}_{Mi}} \right) \\ 
\end{aligned} \right.\\
&\Rightarrow \left\{ \begin{aligned}
  & {{{\dot{\delta }}}_{Mi}}={{\omega }_{b}}\cdot {{{\bar{\omega }}}_{Mi}}={{\omega }_{b}}({{\omega }_{Mi}}-{{\omega }_{g}}) \\ 
 & {{{\dot{\bar{\omega }}}}_{Mi}}={{k}_{pi}}{{\omega }_{ci}}\left( P_{Mi}^\text{ref}-{{P}_{Mi}} \right)-{{\omega }_{ci}}{{{\bar{\omega }}}_{Mi}} \\ 
\end{aligned} \right.
\end{aligned}
\end{equation*}
where $k_{pi}$ is the droop coefficient and $\omega_{ci}$ is the cutoff angular frequency of the low-pass filter of the $i$-th GFM inverter. It can be observed that the two control schemes are mathematically equivalent under the parameter setting of $\frac{1}{J_i}={{k}_{pi}}{{\omega }_{ci}}$ and $\frac{D_i}{J_i}={\omega }_{ci}$. Therefore, for clarity and conciseness, this paper adopts the VSG control framework in Fig. 3(b) without loss of generality.}

Investigation of (\ref{eq:originPLL}) and (\ref{VSGcontrol}) reveals that the explicit derivation of the expressions of $u_{Li}^q$ for GFL inverters and $i_{Mi}^d$ for GFM inverters are required to facilitate the detailed formulation of inverter dynamic models. Accordingly, based on (\ref{eq:Ynewsb}), the voltages of GFL inverters are derived as:
\begin{equation}    \mathbf{u}_{\mathbf{L}}=\mathbf{Y}_{\mathbf{L}}^{-1}\mathbf{i}_{\mathbf{L}}-\mathbf{Y}_{\mathbf{L}}^{-1}\mathbf{Y}_{\mathbf{LM}}{\mathbf{u}_{\mathbf{M}}}-\mathbf{Y}_{\mathbf{L}}^{-1}\mathbf{Y}_{\mathbf{Lg}}{{u}_{g}}.
 \label{eq:uLM}
\end{equation}

\begin{figure*}[!b] 
\setcounter{mytempeqncnt}{\value{equation}}
    \setcounter{equation}{11}
    \hrule 
    \vspace{5pt}  
\vspace{5pt}
\begin{subequations}
    \begin{equation}
    \begin{aligned}
u_{Li}^{q}& ={{\omega }_g}\tilde{\mathbf{L}}_{(i,i)}I_{Li}^{d}+{\tilde{\mathbf{R}}_{(i,i)}}I_{Li}^{q}-{{U}_{g}}\left( \mathbf{K}_{\mathbf{Lg}(i)}^\mathbf{Im}\cos {{\delta }_{Li}}-\mathbf{K}_{\mathbf{Lg}(i)}^\mathbf{Re}\sin {{\delta }_{Li}} \right)+ \\ 
 & {\sum}_{j=1,j\ne i}^{n}{\left( { {A}_{ij}}\cos {{\delta }_{LLij}}-{ {B}_{ij}}\sin {{\delta }_{LLij}} \right)}-{\sum}_{j=1}^{m}{U_{Mj}^{d}\left( \mathbf{K}_{\mathbf{LM}(i,j)}^\mathbf{Im}\cos {{\delta }_{LMij}}-\mathbf{K}_{\mathbf{LM}(i,j)}^\mathbf{Re}\sin {{\delta }_{LMij}} \right)}
    \end{aligned}
\label{eq:uqi}
\end{equation}
\begin{equation}
    \begin{aligned}
 i_{Mi}^{d}&={\tilde{\mathbf{G}}}_{\mathbf{M}(i,i)}U_{Mi}^{d}+{{U}_{g}}({\tilde{\mathbf{G}}}_{\mathbf{Mg}(i)}\cos {{\delta }_{Mi}}+{{{\tilde{\mathbf{B}}}}_{\mathbf{Mg}(i)}}\sin {{\delta }_{Mi}})+{\sum}_{j=1}^{n}{\left( { {a}_{ij}}\cos {{\delta }_{MLij}}+{ {b}_{ij}}\sin {{\delta }_{MLij}} \right)}+\\
&{\sum}_{j=1,j\ne i}^{m}{\left( U_{Mj}^{d}{{{\tilde{\mathbf{G}}}}_{\mathbf{M}(i,j)}}\cos {{\delta }_{MMij}}+U_{Mj}^{d}{{{\tilde{\mathbf{B}}}}_{\mathbf{M}(i,j)}}\sin {{\delta }_{MMij}} \right)}
    \end{aligned}
\label{eq:imi}
\end{equation}
\end{subequations}
\setcounter{equation}{\value{mytempeqncnt}}
\end{figure*}

Substituting (\ref{eq:uLM}) into (\ref{eq:Ynewsb}), the currents of GFM inverters are derived as:
\begin{equation}
\begin{aligned}
\mathbf{i}_{\mathbf{M}}=&\mathbf{Y}_{\mathbf{ML}}\mathbf{Y}_{\mathbf{L}}^{-1}\mathbf{i}_{\mathbf{L}}+(\mathbf{Y}_{\mathbf{M}}-\mathbf{Y}_{\mathbf{ML}}\mathbf{Y}_{\mathbf{L}}^{-1}\mathbf{Y}_{\mathbf{LM}}){\mathbf{u}_{\mathbf{M}}}+\\&(\mathbf{Y}_{\mathbf{Mg}}-\mathbf{Y}_{\mathbf{ML}}\mathbf{Y}_{\mathbf{L}}^{-1}\mathbf{Y}_{\mathbf{Lg}}){{u}_{g}}.
\end{aligned}
\label{iM}
\end{equation}

Based on (\ref{eq:uLM}) and (\ref{iM}), we define:
\begin{subequations}
\begin{equation}
\left\{\begin{aligned}    &\tilde{\mathbf{Z}}=\tilde{\mathbf{R}}+j{\bm{\omega}} \tilde{\mathbf{L}} =\mathbf{Y}_{\mathbf{L}}^{-1}\\
&\mathbf{K}_\mathbf{LM}=\mathbf{K}_\mathbf{LM}^{\mathbf{Re}}+j\mathbf{K}_\mathbf{LM}^{\mathbf{Im}}=\mathbf{Y}_{\mathbf{L}}^{-1}\mathbf{Y}_{\mathbf{LM}} \\
&\mathbf{K}_\mathbf{Lg}=\mathbf{K}_\mathbf{Lg}^{\mathbf{Re}}+j\mathbf{K}_\mathbf{Lg}^{\mathbf{Im}}=\mathbf{Y}_{\mathbf{L}}^{-1}\mathbf{Y}_{\mathbf{Lg}} 
\end{aligned}\right.\
\label{eq:ZK}
\end{equation}
\begin{equation}
\left\{\begin{aligned}
&\mathbf{K}_\mathbf{ML}=\mathbf{K}_\mathbf{ML}^{\mathbf{Re}}+j\mathbf{K}_\mathbf{ML}^{\mathbf{Im}}=\mathbf{Y}_{\mathbf{ML}}\mathbf{Y}_{\mathbf{L}}^{-1}\\
&\tilde{\mathbf{Y}}_\mathbf{M}=\tilde{\mathbf{G}}_\mathbf{M}+j\tilde{\mathbf{B}}_\mathbf{M} =\mathbf{Y}_{\mathbf{M}}-\mathbf{Y}_{\mathbf{ML}}\mathbf{Y}_{\mathbf{L}}^{-1}\mathbf{Y}_{\mathbf{LM}} \\
&\tilde{\mathbf{Y}}_\mathbf{Mg}=\tilde{\mathbf{G}}_\mathbf{Mg}+j\tilde{\mathbf{B}}_\mathbf{Mg} =\mathbf{Y}_{\mathbf{Mg}}-\mathbf{Y}_{\mathbf{ML}}\mathbf{Y}_{\mathbf{L}}^{-1}\mathbf{Y}_{\mathbf{Lg}} 
\end{aligned}\right.\
\label{eq:KY}
\end{equation}
\end{subequations}
where $\tilde{\mathbf{Z}}\in {{\mathbb{R}}^{n\times n}}$, ${\mathbf{K_{LM}}}\in {{\mathbb{R}}^{n\times m}}$, ${\mathbf{K_{Lg}}}\in {{\mathbb{R}}^{n\times 1}}$, ${\mathbf{K_{ML}}}\in {{\mathbb{R}}^{m\times n}}$, $\tilde{\mathbf{Y}}_\mathbf{M}\in {{\mathbb{R}}^{m\times m}}$, and $\tilde{\mathbf{Y}}_\mathbf{Mg}\in {{\mathbb{R}}^{m\times 1}}$. $\tilde{\mathbf{R}}$ and ${\bm{\omega}} \tilde{\mathbf{L}}$ are the real and imaginary parts of $\tilde{\mathbf{Z}}$, $\bm{\omega}=\text{diag}(\omega_{g}, \dots, \omega_{g})\in \mathbb{R}^{n \times n}$. $\tilde{\mathbf{G}}_\mathbf{M}$ and $\tilde{\mathbf{B}}_\mathbf{M}$ are the real and imaginary parts of $\tilde{\mathbf{Y}}_\mathbf{M}$. $\tilde{\mathbf{G}}_\mathbf{Mg}$ and $\tilde{\mathbf{B}}_\mathbf{Mg}$ are the real and imaginary parts of $\tilde{\mathbf{Y}}_\mathbf{Mg}$. The superscript $\mathbf{Re}$ and $\mathbf{Im}$ denote the real and imaginary parts for $\mathbf{K}_\mathbf{LM}$, $\mathbf{K}_\mathbf{Lg}$, $\mathbf{K}_\mathbf{ML}$, respectively.

According to (\ref{eq:ugig})-(\ref{eq:iMuM}) and (\ref{eq:uLM})-(\ref{eq:KY}), the $q$-axis voltage for the $i$-th GFL inverter is expanded in detail as (\ref{eq:uqi}) and the $d$-axis current for the $i$-th GFM inverter is detailed in (\ref{eq:imi}), where:
\addtocounter{equation}{1}
\begin{equation}
\begin{aligned}
&{{\delta }_{LLij}}={{\theta }_{Li}}-{{\theta }_{Lj}}, {{\delta }_{LMij}}={{\theta }_{Li}}-{{\theta }_{Mj}}, \\
&{{\delta }_{MMij}}={{\theta }_{Mi}}-{{\theta }_{Mj}}, {{\delta }_{MLij}}={{\theta }_{Mi}}-{{\theta }_{Lj}},
\end{aligned} 
\end{equation}
and:
\begin{subequations}
\begin{equation}
    \left\{ \begin{aligned}
  & {{A}_{ij}}={\tilde{\mathbf{R}}}_{(i,j)}I_{Lj}^{q}+{{\omega }_{g}}{\tilde{\mathbf{L}}}_{(i,j)}I_{Lj}^{d} \\ 
 & {{B}_{ij}}={\tilde{\mathbf{R}}}_{(i,j)}I_{Lj}^{d}-{{\omega }_{g}}{\tilde{\mathbf{L}}}_{(i,j)}I_{Lj}^{q} 
\end{aligned} \right.
\label{eq:AB}
\end{equation}
\begin{equation}
    \left\{ \begin{aligned}
  & {{a}_{ij}}=\mathbf{K}_{\mathbf{ML}(i,j)}^\mathbf{Re}I_{Lj}^{d}-\mathbf{K}_{\mathbf{ML}(i,j)}^\mathbf{Im}I_{Lj}^{q} \\ 
 & {{b}_{ij}}=\mathbf{K}_{\mathbf{ML}(i,j)}^\mathbf{Im}I_{Lj}^{d}+\mathbf{K}_{\mathbf{ML}(i,j)}^\mathbf{Re}I_{Lj}^{q}. 
\end{aligned} \right.
\label{eq:ab}
\end{equation}
\end{subequations}

Substituting (\ref{eq:uqi}) into (\ref{eq:originPLL}) yields:
\begin{equation}
\begin{aligned}
  & {{{\dot{\bar{\omega }}}}_{Li}}=K_{\text{I}i}^\text{PLL}\left( {{g}_{Li}}-\sum\limits_{j=1,j\ne i}^{n}{g_{LLij} }-\sum\limits_{j=1}^{m}{g_{LMij} } \right) \\ 
 & -K_{\text{P}i}^\text{PLL}{{\omega }_{b}}\left( 
 \begin{aligned}
      &{{d}_{Li}}\cdot{\bar{\omega }}_{Li}+\sum\limits_{j=1,j\ne i}^{n}{{{d}_{LLij}}\cdot{\bar{\omega }}_{LLij}}+\\
      &\sum\limits_{j=1}^{m}{{{d}_{LMij}}\cdot{\bar{\omega }}_{LMij}}
 \end{aligned}
 \right) 
 \end{aligned}
 \label{eq:newPLL}
\end{equation}
where

\begin{subequations}
\begin{equation}
\left\{
\begin{aligned}
  & 
  \begin{aligned}
      {{g}_{Li}}=&{{\omega }_{g}}{{{\tilde{\mathbf{L}}}}_{(i,i)}}I_{Li}^{d}+{{{\tilde{\mathbf{R}}}}_{(i,i)}}I_{Li}^{q}-\\
      &{{U}_{g}}\left( \mathbf{K}_{\mathbf{Lg}(i)}^\mathbf{Im}\cos {{\delta }_{Li}}-\mathbf{K}_{\mathbf{Lg}(i)}^\mathbf{Re}\sin {{\delta }_{Li}} \right) 
  \end{aligned}
  \\ 
 & g_{LLij} =\left( { {B}_{ij}}\sin {{\delta }_{LLij}}-{ {A}_{ij}}\cos {{\delta }_{LLij}} \right) \\ 
 & g_{LMij} ={{U}_{Mj}}\left( \mathbf{K}_{\mathbf{LM}(i,j)}^\mathbf{Im}\cos {{\delta }_{LMij}}-\mathbf{K}_{\mathbf{LM}(i,j)}^\mathbf{Re}\sin {{\delta }_{LMij}} \right) 
 \end{aligned} \right.\label{GFL_g}
\end{equation}

\begin{equation}
\left\{\begin{aligned}
  & {{d}_{Li}}=-{{U}_{g}}\left( \mathbf{K}_{\mathbf{Lg}(i)}^\mathbf{Im}\sin {{\delta }_{Li}}+\mathbf{K}_{\mathbf{Lg}(i)}^\mathbf{Re}\cos {{\delta }_{Li}} \right) \\ 
 & {{d}_{LLij}}={ {A}_{ij}}\sin {{\delta }_{LLij}}+{ {B}_{ij}}\cos {{\delta }_{LLij}} \\ 
 & {{d}_{LMij}}=-{{U}_{Mj}}\left( \mathbf{K}_{\mathbf{LM}(i,j)}^\mathbf{Im}\sin {{\delta }_{LMij}}+\mathbf{K}_{\mathbf{LM}(i,j)}^\mathbf{Re}\cos {{\delta }_{LMij}} \right). \\ 
\end{aligned}\right.
\label{GFL_d}
\end{equation} 
\end{subequations}

In (\ref{GFL_g}) and (\ref{GFL_d}), $g_{Li}$ and ${d_{Li}}$ denote the self-restoring force and self-damping of the $i$-th GFL inverter, respectively. $g_{LLij}$ and $d_{LLij}$ denote the effect of the $j$-th GFL inverter's coupling effect on the $i$-th GFL inverter's restoring force and damping, respectively. Similarly, $g_{LMij}$ and $d_{LMij}$ denote the effect of the $j$-th GFM inverter's coupling effect on the $i$-th GFL inverter's restoring force and damping, respectively. 

Substituting (\ref{eq:imi}) into (\ref{VSGcontrol}) yields:
\begin{equation}
\begin{aligned}
{{\dot{\bar{\omega }}}_{Mi}}=&\frac{1}{{{J}_{i}}}\left( {{g}_{Mi}}-\sum\limits_{j=1,j\ne i}^{m}{g_{MMij} }-\sum\limits_{j=1}^{n}{g_{MLij} } \right)-\\
&\frac{D_i}{{{J}_{i}}}\cdot{{\bar{\omega }}_{Mi}}
\end{aligned}\label{newVSG}
\end{equation}
where
\begin{equation}
\left\{\begin{aligned}
  & \begin{aligned}
      g_{Mi}=&P_{Mi}^\mathrm{ref}-{\tilde{\mathbf{G}}_{\mathbf{M}(i,i)}}{{\left( U_{Mi}^{d} \right)}^{2}}- \\ 
 & U_{Mi}^{d}{{U}_{g}}({\tilde{\mathbf{G}}_{\mathbf{Mg}(i)}}\cos {{\delta }_{Mi}}+{\tilde{\mathbf{B}}_{\mathbf{Mg}(i)}}\sin {{\delta }_{Mi}})
  \end{aligned}\\ 
 & g_{MMij} =U_{Mi}^{d}U_{Mj}^{d}\left(
 \begin{aligned}
      &{\tilde{\mathbf{G}}_{\mathbf{M}(i,j)}}\cos {{\delta }_{MMij}}+\\
      &{\tilde{\mathbf{B}}_{\mathbf{M}(i,j)}}\sin {{\delta }_{MMij}}
 \end{aligned}
 \right) \\ 
 & g_{MLij} =U_{Mi}^{d}\left( {{a}_{ij}}\cos {{\delta }_{MLij}}+{{b}_{ij}}\sin {{\delta }_{MLij}} \right).
\end{aligned}\right.\label{gM}
\end{equation}

Analogous to GFL inverter dynamics, in (\ref{gM}), $g_{Mi}$ denotes the self-restoring force of the $i$-th GFM inverter. $g_{MMij}$ and $g_{MLij}$ characterize the effect of the $j$-th GFM inverter and the $j$-th GFL inverter's coupling effect on the $i$-th GFM inverter's restoring force, respectively.

By comparing the GFL and GFM inverter dynamics in (\ref{eq:newPLL}) and (\ref{newVSG}), a clear duality between the two can be observed. The equivalent inertia for GFL inverter is represented by $1/K_{\text{I}i}^\text{PLL}$ and for GFM inverter is $J_i$. The damping coefficient for GFM inverter is given by $D_i$. While for GFL inverters, investigation of (\ref{GFL_d}) reveals that their equivalent damping varies dynamically with $\delta_{Li}$, $\delta_{LLij}$ and $\delta_{LMij}$. 

Rearranging (\ref{eq:originPLL}), (\ref{VSGcontrol}), (\ref{eq:newPLL}) and (\ref{newVSG}), system dynamics is expressed in vector form as:
\begin{equation}
\left\{ \begin{aligned}
  & {{{\dot{\bm{\delta} }}}_{L}}={{\omega }_{b}}{{{\bar{\bm{\omega} }}}_{L}} \\   
& {{{\dot{\bm{\delta} }}}_{M}}={{\omega }_{b}}{{{\bar{\bm{\omega} }}}_{M}} \\  
 & \dot{{{\bar{\bm{\omega} }}}}_{L}=\mathbf{K}_{\mathbf{I}}^{\mathbf{PLL}}\left({{\mathbf{G}}_{\mathbf{L}}}-\mathbf{G}_{\mathbf{LL}}-\mathbf{G}_{\mathbf{LM}}\right)-\\
   & \quad \quad \ \ \omega_b \mathbf{K}_{\mathbf{P}}^{\mathbf{PLL}}\left( {{\mathbf{D}}_{\mathbf{L}}}{{{\bar{\bm{\omega}}}}_{L}} - {{\mathbf{D}}_{\mathbf{LM}}}{{{\bar{\bm{\omega}}}}_{M}} \right)\\  
& \dot{{{\bar{\bm{\omega} }}}}_{M}=\mathbf{J}^{-1}\left({{\mathbf{G}}_{\mathbf{M}}}-\mathbf{G}_{\mathbf{MM}}-\mathbf{G}_{\mathbf{ML}}\right)-\mathbf{J}^{-1}\mathbf{D}_{\mathbf{M}}{\bar{\bm{\omega}}}_{M}
\end{aligned} \right.\label{vec}
\end{equation}
\begin{figure*}[b]  
    \hrule 
    \vspace{5pt}
\begin{subequations}  
\begin{equation}
   \mathbf{G}_{\mathbf{L}}={{\left[ {{g}_{L1}},\cdots ,{{g}_{Ln}} \right]}^{\top}},\quad \mathbf{G}_{\mathbf{LL}}={{\left[ \sum\limits_{j=2}^{n}{g_{LL1j} },\cdots ,\sum\limits_{j=1}^{n-1}{g_{LLnj} } \right]}^{\top}},\quad \mathbf{G}_{\mathbf{LM}}={{\left[ \sum\limits_{j=1}^{m}{g_{LM1j} },\cdots ,\sum\limits_{j=1}^{m}{g_{LMnj} } \right]}^{\top}} 
  \label{allGL}
\end{equation}
\begin{equation}
 \mathbf{G}_{\mathbf{M}}={{\left[ {{g}_{M1}},\cdots ,{{g}_{Mm}} \right]}^{\top}},\quad \mathbf{G}_{\mathbf{MM}}={{\left[ \sum\limits_{j=2}^{m}{g_{MM1j} },\cdots ,\sum\limits_{j=1}^{m-1}{g_{MMmj} } \right]}^{\top}},\quad \mathbf{G}_{\mathbf{LM}}={{\left[ \sum\limits_{j=1}^{n}{g_{ML1j} },\cdots ,\sum\limits_{j=1}^{n}{g_{MLmj} } \right]}^{\top}}
  \label{allGM}
\end{equation}
\begin{equation}
    \mathbf{D}_{\mathbf{L}}=\left[ \begin{matrix}
   {{d}_{L1}}+\sum\limits_{j=2}^{n}{d_{LL1j} }+\sum\limits_{j=1}^{m}{d_{LM1j} } & -d_{LL12}  & \cdots  & -d_{LL1n}   \\
   -d_{LL21}  & {{d}_{L2}}+\sum\limits_{j=1,j\ne 2}^{n}{d_{LL2j} }+\sum\limits_{j=1}^{m}{d_{LM2j} } & \cdots  & -d_{LL2n}   \\
   \vdots  & \vdots  & \ddots  & \vdots   \\
   -d_{LLn1}  & -d_{LLn2}  & \cdots  & {{d}_{Ln}}+\sum\limits_{j=1}^{n-1}{d_{LLnj} }+\sum\limits_{j=1}^{m}{d_{LMnj} }  \\
\end{matrix} \right]
\label{DL}
\end{equation}
\end{subequations}
\end{figure*}
where $\bm{\delta}_L={{[\delta_{L1},...,\delta_{Ln}]}^\top}$, $\bar{\bm{\omega}}_L={[\bar{\omega }_{L1},...,{\bar{\omega }}_{Ln}]}^\top$, $\bm{\delta}_M={{[\delta_{M1},...,\delta_{Mm}]}^\top}$ and $\bar{\bm{\omega}}_M={[\bar{\omega }_{M1},...,{\bar{\omega }}_{Mm}]}^\top$. The control parameters are denoted by $\mathbf{K}_{\mathbf{I}}^{\mathbf{PLL}}=\mathrm{diag}(K_{\text{I}1}^\text{PLL}, \dots, K_{\text{I}n}^\text{PLL})$, $\mathbf{K}_{\mathbf{P}}^{\mathbf{PLL}}=\mathrm{diag}(K_{\text{P}1}^\text{PLL}, \dots, K_{\text{P}n}^\text{PLL})$, $\mathbf{J}=\mathrm{diag}(J_1, \dots , J_m)$ and $\mathbf{D}_{\mathbf{M}}=\mathrm{diag}(D_1, \dots , D_m)$. $\mathbf{D}_{\mathbf{LM}}(i,j) = d_{LMij}$ where $ i=1,\dots,n$ and $ j=1,\dots,m$. The expressions of $\mathbf{G}_{\mathbf{L}}$, $\mathbf{G}_{\mathbf{LL}}$, $\mathbf{G}_{\mathbf{LM}}$, $\mathbf{G}_{\mathbf{M}}$, $\mathbf{G}_{\mathbf{MM}}$ , $\mathbf{G}_{\mathbf{ML}}$ and $\mathbf{D}_{\mathbf{L}}$ are detailed in (\ref{allGL})-(\ref{DL}). At the SEP $({\bm{\delta}_L^s},{\bm{\delta}_M^s},{\bar{\bm{\omega}}_L^s},{\bar{\bm{\omega}}_M^s})=({\bm{\delta}_L^s},{\bm{\delta}_M^s},{\mathbf{0}},{\mathbf{0}})$, the right-hand sides of (\ref{vec}) are equal to $\mathbf{0}$ and the Jacobian matrix of (\ref{vec}) at $({\bm{\delta}_L^s},{\bm{\delta}_M^s},{\mathbf{0}},{\mathbf{0}})$ is negative definite. 
\vspace{-0.3cm}
\section{Transient Stability Analysis}
\subsection{Local Energy Function Candidate}
To analytically conduct transient stability analysis for the studied mixed GFM–GFL-inverter system, the energy function method is employed in this paper. \rv{For a single GFL inverter, its Lyapunov function has been well established in prior studies. For systems with multiple GFM inverters, the system energy function can be constructed in a manner similar to that for conventional multi-machine synchronous generator systems, where the lossy term related to network transfer conductance can typically be ignored. However, for systems with multiple GFL inverters, the lossy term introduced by $A_{ij}\cos\delta_{LLij}$ in (\ref{GFL_g}) poses significant challenges for constructing a system-wide energy function.} Nevertheless, a local energy function candidate is proposed herein, and it is demonstrated in the following subsection that this function is decreasing within a certain region $\Omega$, thereby validating its applicability.

The equivalent kinetic energy and potential energy for the GFM and GFL inverters are defined as follows:
\begin{equation}
    {{V}_{KL}}=\sum\limits_{i=1}^{n}{\frac{{{\omega }_{b}}}{2K_{Ii}^{PLL}}\bar{\omega }_{Li}^{2}},\quad{{V}_{KM}}=\sum\limits_{i=1}^{m}{\frac{{{\omega }_{b}}{{J}_{i}}}{2}\bar{\omega }_{Mi}^{2}}
    \label{KineticEnergy}
\end{equation} 

\begin{equation}
    \begin{aligned}
   {{V}_{PL}}&=\sum\limits_{i=1}^{n}{\left[ \begin{aligned}
  & {{U}_{g}}\left[ \begin{aligned}
      &\mathbf{K}_{\mathbf{Lg}(i)}^\mathbf{Re}(\cos {{\delta }_{Li}}-\cos \delta _{Li}^{s})+\\
      &\mathbf{K}_{\mathbf{Lg}(i)}^\mathbf{Im}(\sin {{\delta }_{Li}}-\sin \delta _{Li}^{s})
  \end{aligned} \right]- \\ 
 & ({{\omega }_{g}}{\tilde{\mathbf{L}}_{(i,i)}}I_{Li}^{d}+{\tilde{\mathbf{R}}_{(i,i)}}I_{Li}^{q})\cdot ({{\delta }_{Li}}-{{\delta }_{Li}^s}) \\ 
\end{aligned} \right]}+ \\ 
 & \sum\limits_{i=1}^{n}{\sum\limits_{j=1,j\ne i}^{n}{\left[
 \begin{aligned}
     &-{ {B}_{ij}}(\cos \delta _{LLij}^{*}-\cos \delta _{LLij}^{s})-\\&{ {A}_{ij}}(\sin \delta _{LLij}^{*}-\sin \delta _{LLij}^{s})
 \end{aligned} \right]}}+ \\ 
 & \sum\limits_{i=1}^{n}{\sum\limits_{j=1}^{m}{U_{Mj}^{d}\left[ 
 \begin{aligned}
&\mathbf{K}_{\mathbf{LM}(i,j)}^\mathbf{Re}(\cos {{\delta }_{LMij}}-\cos \delta _{LMij}^{s})+\\
&\mathbf{K}_{\mathbf{LM}(i,j)}^\mathbf{Im}(\sin {{\delta }_{LMij}}-\sin \delta _{LMij}^{s})    
 \end{aligned}\right]}} \\ 
\end{aligned} \label{VPL}
\end{equation}

\rv{\begin{equation}
\begin{aligned}
   {{V}_{PM}}&=\sum\limits_{i=1}^{m}\left[ \begin{aligned}
  & U_{Mi}^{d}{{U}_{g}}\left[ \begin{aligned}
  & \tilde{\mathbf{G}}_{\mathbf{Mg}(i)}(\sin {{\delta }_{Mi}}-\sin \delta _{Mi}^{s})- \\ 
 & \tilde{\mathbf{B}}_{\mathbf{Mg}(i)}(\cos {{\delta }_{Mi}}-\cos \delta _{Mi}^{s}) \\ 
\end{aligned} \right]+ \\ 
 & \left( (U_{Mi}^{d})^2{{{\tilde{\mathbf{G}}}}_{\mathbf{M}(i,i)}}-P_{Mi}^\mathrm{ref} \right)\cdot ({{\delta }_{Mi}}-\delta _{Mi}^{s}) \\ 
\end{aligned} \right]- \\ 
 & \sum\limits_{i=1}^{m-1}{\sum\limits_{j=i+1}^{m}{U_{Mi}^{d}U_{Mj}^{d}\left[ \begin{aligned}
  & \tilde{\mathbf{G}}_{\mathbf{M}(i,j)}(\sin {{\delta }_{MMij}}-\sin \delta _{MMij}^{s})+ \\ 
 & \tilde{\mathbf{B}}_{\mathbf{M}(i,j)}(\cos {{\delta }_{MMij}}-\cos \delta _{MMij}^{s}) \\ 
\end{aligned} \right]}}+ \\ 
 &\sum\limits_{i=1}^{m} \sum\limits_{j=1}^{n}{U_{Mi}^{d}\left[ \begin{aligned}
  & -{ {b}_{ij}}(\cos {{\delta }_{MLij}}-\cos \delta _{MLij}^{s})+ \\ 
 & { {a}_{ij}}(\sin {{\delta }_{MLij}}-\delta _{MLij}^{s}) \\ 
\end{aligned} \right]} \\ 
\end{aligned}\label{PotentialEnergy}
\end{equation}}
where $\delta_{LLij}^s=\delta_{Li}^s-\delta_{Lj}^s$, $\delta_{LMij}^s=\delta_{Li}^s-\delta_{Mj}^s$, $\delta_{MMij}^s=\delta_{Mi}^s-\delta_{Mj}^s$ and $\delta_{MLij}^s=\delta_{Mi}^s-\delta_{Lj}^s$. In (\ref{VPL}), $\delta_{LLij}^*=\delta_{Li}-\delta_{Lj}^s$. $V_{KL}$ and $V_{KM}$ denote the kinetic energy for the GFL and GFM inverters in the system, respectively. $V_{PL}$ and $V_{PM}$ represent the potential energy for the GFL and GFM inverters, respectively.

\rv{Specifically, the first term and the third terms of $V_{PL}$ are obtained by taking the first integrals of $g_{Li}$ and $g_{LMij}$ in (16a), respectively. In accordance with (15), a factor of $-1$ is multiped to the first term, such that these terms cancel out in the derivation of ${{{\dot{V}}}_{PL}}+{{{\dot{V}}}_{KL}}$ in the subsequent discussion. The second term of $V_{PL}$ is constructed via the integration of $\delta_{LLij}^*=\delta_{Li}-\delta_{Lj}^s$. This design choice specifically addresses the non-negligible term $A_{ij}\cos\delta_{LLij}$ in (16a) introduced by GFL–GFL coupling. As for GFM inverters' potential energy, all three terms in $V_{PM}$ are obtained by taking the first integrals of $g_{Mi}$, $g_{MMij}$ and $g_{MLij}$ in (18), respectively. And similarly, a factor of $-1$ is also multiped to the first term based on expression of (17).}

\rv{When constructing $V_{PL}$ to account for GFL–GFL coupling, the shifted angle variable $\delta_{LLij}^*$ is adopted. Consequently, the second term in (22) is expressed as a double summation over $i=1,\dots,n$ and $j=1,\dots,n$ with $j\ne i$. For the construction of $V_{PM}$ associated with GFM–GFM coupling, due to the symmetric of $\tilde{\mathbf{G}}_{\mathbf{M}}$ and $\tilde{\mathbf{B}}_{\mathbf{M}}$, the second term in (23) is formulated using the reduced double summation $i=1,\dots,m-1$ and $j=i+1,\dots,m$ to avoid double counting of symmetric terms. }

Combining (\ref{KineticEnergy})-(\ref{PotentialEnergy}), the local energy function candidate for the mixed GFM-GFL-inverter system is constructed as:
\begin{equation}
V({\bm{\delta}_L},{\bm{\delta}_M},{\bar{\bm{\omega}}_L},{\bar{\bm{\omega}}_M})={{V}_{KL}}+{{V}_{PL}}+{{V}_{KM}}+{{V}_{PM}}.
\label{Ly}
\end{equation}

At the SEP $({\bm{\delta}_L^s},{\bm{\delta}_M^s},{\mathbf{0}},{\mathbf{0}})$, $V({\bm{\delta}_L^s},{\bm{\delta}_M^s},{\mathbf{0}},{\mathbf{0}})=0$. \rv{It can be seen from (\ref{KineticEnergy})-(\ref{PotentialEnergy}) that a bounded value of $V$ for time $t \in \mathbb{R}^+$ implies that the system sates $({\bm{\delta}_L},{\bm{\delta}_M},{\bar{\bm{\omega}}_L},{\bar{\bm{\omega}}_M})$ are also bounded. Regarding the frequency states, the kinetic energy terms $V_{KL}$ and $V_{KM}$ in (21) are constructed as sums of squared frequency deviations weighted by positive coefficients. If the norm $\| \bar{\bm{\omega}}_L \|$ or $\| \bar{\bm{\omega}}_M \|$ tends to infinity, $V_{KL}+V_{KM}$ would tend to infinity. Therefore, a bounded total energy $V$ strictly implies that the frequency states $\bar{\bm{\omega}}$ must be bounded. Regarding the angle states, the potential energy terms $V_{PL}$ and $V_{PM}$ consist of two distinct types of mathematical components: trigonometric terms involving sine and cosine functions, and linear terms proportional to the angle deviations $(\delta - \delta^s)$, shown as $({{\omega }_{g}}{\tilde{\mathbf{L}}_{(i,i)}}I_{Li}^{d}+{\tilde{\mathbf{R}}_{(i,i)}}I_{Li}^{q})\cdot ({{\delta }_{Li}}-{{\delta }_{Li}^s})$ in (22) and as $\left( (U_{Mi}^{d})^2{{{\tilde{\mathbf{G}}}}_{\mathbf{M}(i,i)}}-P_{Mi}^\mathrm{ref} \right)\cdot ({{\delta }_{Mi}}-\delta _{Mi}^{s})$ in (23), respectively. It is observed that while the trigonometric terms are globally bounded regardless of the state values, the linear terms are inherently unbounded. As the magnitude of the angle deviations $\| \bm{\delta} \|$ increases towards infinity, the unbounded linear terms will inevitably dominate the bounded trigonometric terms. Consequently, the magnitude of the potential energy $|V_P|$ would tend to infinity (i.e., $\| \bm{\delta} \| \to \infty \implies |V_P| \to \infty$). Therefore, since both $V_K$ and $V_P$ would diverge if their respective states diverged, the condition of finite system energy $V$ ensures that both the frequency states $\bar{\bm{\omega}}$ and the angle states $\bm{\delta}$ remain bounded for all positive time.}

\subsection{Critical Energy and Stability Index}
For a global energy function, it is required that the derivative of the energy function along any system trajectory be nonpositive, i.e., $\dot{V}\le 0$. 
\rv{Such condition makes it infeasible to incorporate GFL interactions, since their coupling can create regions where $\dot{V} \le 0$ no longer holds. In contrast, for the proposed local energy function, it suffices for this condition to hold within a specified region where the combined equivalent damping of GFL and GFM inverters remains positive, thereby systematically incorporating the interaction effects of both GFL and GFM inverters.}

Taking the time derivative of (\ref{Ly}) along the system trajectory described by (\ref{vec}) for the GFL and GFM inverters, respectively, yields:
\begin{equation}
\begin{aligned}
   &{{{\dot{V}}}_{PL}}+{{{\dot{V}}}_{KL}}=\sum\limits_{i=1}^{n}{\left[ \frac{\partial {{V}_{PL}}}{\partial {{\delta }_{Li}}}\cdot {{{\dot{\delta }}}_{Li}}+\frac{{{V}_{KL}}}{\partial {{{\bar{\omega }}}_{Li}}}\cdot {{{\dot{\bar{\omega }}}}_{Li}} \right]} \\ 
 & ={{\omega }_{b}}\left( \bm{\bar{\omega}_L}^\top \mathbf{\Gamma} -\bm{\bar{\omega}_L}^\top{\mathbf{\Lambda_L}}\mathbf{{D}_{L}}\bm{\bar{\omega}_L}+\bm{\bar{\omega}_L}^\top{\mathbf{\Lambda_L}}\mathbf{{D}_{LM}}\bm{\bar{\omega}_M} \right)  
\end{aligned}\label{dVL}
\end{equation}

\begin{equation}
\begin{aligned}
  & {{{\dot{V}}}_{PM}}+{{{\dot{V}}}_{KM}}=\sum\limits_{i=1}^{m}{\left[ \frac{\partial {{V}_{PM}}}{\partial {{\delta }_{Mi}}}\cdot {{{\dot{\delta }}}_{Mi}}+\frac{{V_{KM}}}{\partial {{{\bar{\omega }}}_{Mi}}}\cdot {{{\dot{\bar{\omega }}}}_{Mi}} \right]} \\ 
 & =-2{{\omega }_{b}}\sum\limits_{i=1}^{m-1}{\sum\limits_{j=i+1}^{m}{U_{Mi}^{d}U_{Mj}^{d}\left( {{{\tilde{\mathbf{G}}}}_{\mathbf{M}(i,j)}}\cos {{\delta }_{MMij}}\cdot {{{\bar{\omega }}}_{Mi}} \right)}}\\
 &\quad -{{\omega }_{b}}\bar{\bm{\omega}}_{M}^\top{{\mathbf{D}}_{\mathbf{M}}}{{{\bar{\bm{\omega}}}}_{M}} \\ 
\end{aligned}
\label{dVM}
\end{equation}
where
\begin{equation}
\begin{aligned}
\mathbf{\Gamma}(i) & = \sum\limits_{j=1,j\neq i}^{n} 2\sqrt{A_{ij}^{2}+B_{ij}^{2}} \sin \left( \frac{\delta_{Lj}-\delta_{Lj}^{s}}{2} \right) \cdot \\
& \sin \left( \delta_{Li}-\frac{\delta_{Lj}^{s}+\delta_{Lj}}{2}+\varphi_{ij} \right), \quad \mathbf{\Gamma} \in \mathbb{R}^{n\times1}
\end{aligned}
\label{eq:Gamma}
\end{equation}
\begin{equation}
    \varphi_{ij}=\arctan(B_{ij}/A_{ij})
    \label{eq:Gamma2}
\end{equation}
\begin{equation}
    \mathbf{\Lambda}_\mathbf{L}={{\omega }_{b}}{{(\mathbf{K}_{\mathbf{I}}^{\mathbf{PLL}})}^{-1}}\mathbf{K}_{\mathbf{P}}^{\mathbf{PLL}}\label{Lambda_L}.
\end{equation}


\rv{Considering that high-voltage transmission networks typically exhibit a high reactance-to-resistance ratio ($X \gg R$), the transfer conductances are significantly smaller than the susceptances. Consequently, comparing (\ref{dVL}) and (\ref{dVM}), it can be observed that the lossy term introduced by $\tilde{\mathbf{G}}_{\mathbf{M}(i,j)}$ in (\ref{dVM}) becomes negligible \rv{\cite{Chiang,Fouad,Wang_Zhi}}.} As a result, the dissipative effect in (\ref{dVM}) vanishes, and the combined derivative terms satisfy $(\dot{V}_{PM} + \dot{V}_{KM}) < 0$. \rv{Since the load characteristics are predominantly retained within the self-admittance terms, the omission of the minor load component in $\tilde{\mathbf{G}}_{\mathbf{M}(i,j)}$ is physically equivalent to neglecting a resistive dissipative element. This simplification inherently contributes to the conservative nature of the proposed energy function framework.  }

In contrast, for the GFL inverters, the lossy term $\bar{\bm{\omega}}_{L}^{T}\mathbf{\Gamma}$ in (\ref{dVL}) cannot be neglected. This term arises due to the interaction of the current source behavior of GFL inverters with the transfer impedances of the network as seen in (\ref{eq:AB}), which include both resistance and reactance. Since the GFL inverter controls current rather than voltage, it does not suppress the resistive dissipation introduced by the network. Therefore, the lossy term $\bar{\bm{\omega}}_{L}^{T}\mathbf{\Gamma}$ persists and  requires further investigation in the transient stability analysis.

Accordingly, by neglecting the lossy term in (\ref{dVM}), the combination of (\ref{dVL}) and (\ref{dVM}) yields:
\begin{equation}
\dot{V}={{\omega }_{b}}\left( \begin{aligned}
    &\bar{\bm{\omega}}_{L}^{\top}\mathbf{\Gamma} -\bar{\bm{\omega}}_{L}^{\top}\mathbf{\Lambda}_\mathbf{L}{\mathbf{D}_\mathbf{L}}{{{\bar{\bm{\omega}}}}_{L}}+\bar{\bm{\omega}}_{L}^{\top}\mathbf{\Lambda}_\mathbf{L}{{\mathbf{D}}_{\mathbf{LM}}}{{{\bar{\bm{\omega}}}}_{M}} \\
    &- \bar{\bm{\omega}}_{M}^{\top}{{\mathbf{D}}_{\mathbf{M}}}{{{\bar{\bm{\omega}}}}_{M}}
    \end{aligned}\right).
\label{dVs}
\end{equation}

Based on (\ref{dVs}), a scalar function with respect to $\bar{\bm{\omega}}_{M}$ is defined as:
\begin{equation}
\begin{aligned}
    f(\bar{\bm{\omega}}_{M})=
    &-\bar{\bm{\omega}}_{M}^{\top}{{\mathbf{D}}_{\mathbf{M}}}{{{\bar{\bm{\omega}}}}_{M}}+\bar{\bm{\omega}}_{L}^{\top}\mathbf{\Lambda}_\mathbf{L}{{\mathbf{D}}_{\mathbf{LM}}}{{{\bar{\bm{\omega}}}}_{M}}+\\
    &\bar{\bm{\omega}}_{L}^{\top}\mathbf{\Gamma} -\bar{\bm{\omega}}_{L}^{\top}\mathbf{\Lambda}_\mathbf{L}{\mathbf{D}_\mathbf{L}}{{{\bar{\bm{\omega}}}}_{L}}\\
\end{aligned} \label{fwM}
\end{equation}
where $f(\bar{\bm{\omega}}_{M})<0$ implies that $\dot{V}<0$ holds. 

\rv{Differentiating $f(\bar{\bm{\omega}}_{M})$ with respect to $\bar{\bm{\omega}}_{M}$ yields:
\begin{equation*}
\begin{aligned}
&\frac{\partial f}{\partial \bar{\bm{\omega}}_M} = -2\mathbf{D_M} \bar{\bm{\omega}}_M + \mathbf{D}_\mathbf{LM}^\top \mathbf{\Lambda_L} \bar{\bm{\omega}}_L= \mathbf{0}\\
    &\Rightarrow \bar{\bm{\omega}}_M^* = \frac{1}{2} \mathbf{D}_\mathbf{M}^{-1} \mathbf{D}_\mathbf{LM}^\top \mathbf{\Lambda_L} \bar{\bm{\omega}}_L   
\end{aligned}
\end{equation*}}

With the damping coefficients of GFM inverters being positive (${{\mathbf{D}}_{\mathbf{M}}}>\mathbf{0}$), the maximum value of (\ref{fwM}) with respect to $\bar{\bm{\omega}}_M$ is derived as:\rv{
\begin{equation}
  {{f}_{\max }}=f({{\bar{\bm{\omega}}}_{M}^*})={{\mathbf{\Gamma} }^{\top}}{{\bar{\bm{\omega}}}_{L}}+\bar{\bm{\omega}}_{L}^{\top}\mathbf{Q}{{\bar{\bm{\omega}}}_{L}}\label{fmax}
\end{equation}}
where
\begin{equation}
    \mathbf{Q}= \frac{\mathbf{\Lambda}_\mathbf{L}{{\mathbf{D}}_{\mathbf{LM}}}{\mathbf{D}}_{\mathbf{M}}^{-1}{{(\mathbf{\Lambda}_\mathbf{L}{\mathbf{D}}_{\mathbf{LM}})}^\top}}{4}-\mathbf{\Lambda}_\mathbf{L}{\mathbf{D}}_{\mathbf{L}}.
\end{equation}

For $f(\bar{\bm{\omega}}_{M})<0$ to hold for all non-zero vectors $\bar{\bm{\omega}}_\mathbf{L}$ and $\bar{\bm{\omega}}_\mathbf{M}$, ${{f}_{\max }}({{\bar{\bm{\omega}}}_{M}})<0$ should be satisfied. Considering that $\mathbf{Q}=\mathbf{Q_s}+\mathbf{Q_a}$ where $\mathbf{Q_s}=(\mathbf{Q}+\mathbf{Q}^\top)/2$ and $\mathbf{Q_a}=(\mathbf{Q}-\mathbf{Q}^\top)/2$, then:
\begin{equation}
\bar{\bm{\omega}}_{L}^{\top}\mathbf{Q}{{\bar{\bm{\omega}}}_{L}}=\bar{\bm{\omega}}_{L}^{\top}\mathbf{Q_s}{{\bar{\bm{\omega}}}_{L}} \quad (\bar{\bm{\omega}}_{L}^{\top}\mathbf{Q_a}{{\bar{\bm{\omega}}}_{L}}=\mathbf{0}).
\end{equation}

Accordingly, (\ref{fmax}) can be rewritten as:
\begin{equation}
    {{\left( {\bar{\bm{\omega}}_L}+\frac{1}{2}\mathbf{Q_s^{-1}}\mathbf{\Gamma}  \right)}^{\top}}\mathbf{Q_s}\left( {\bar{\bm{\omega}}_L}+\frac{1}{2}\mathbf{Q_s^{-1}}\mathbf{\Gamma}  \right)-\frac{1}{4}{{\mathbf{\Gamma} }^{\top}}\mathbf{Q_s^{-1}}\mathbf{\Gamma} <0.
    \label{condi}
\end{equation}

Under stable operation, ${\bar{\bm{\omega}}_L}=\mathbf{0}$ and (35) evaluates to zero, confirming that $\dot{V}=0$ at the SEP. For the case where ${\bar{\bm{\omega}}_L} \ne \mathbf{0}$, the condition relies on the negative definiteness of matrix $\mathbf{Q_s}$. If $\mathbf{Q_s}$ is negative definite, the first term in (35) is strictly negative. Conversely, the second term, $-\frac{1}{4}{{\mathbf{\Gamma}}^{\top}}\mathbf{Q_s^{-1}}\mathbf{\Gamma}$, becomes positive given the negative definiteness of $\mathbf{Q_s^{-1}}$ and the coefficient $-1/4$. However, as shown in (27), the residual vector $\mathbf{\Gamma}$ is composed of bounded sinusoidal functions and is independent of the frequency deviation ${\bar{\bm{\omega}}_L}$. \rvb{Consequently, when the magnitude of the frequency deviation exceeds a threshold, specifically $\|\bar{\bm{\omega}}_L\|>r$, the negative contribution of the quadratic first term dominates the bounded positive second term, ensuring the inequality in (35) holds. Physically, this means that once the post-fault frequency deviation grows beyond the bound $r$, the system energy strictly decays ($\dot V<0$), driving the trajectory back toward a bounded residual neighborhood of the SEP. This property guarantees that the post-fault trajectories are ultimately confined within a bounded region $\Omega$ around the SEP, a behavior referred to as Lagrange stability~\cite{El_abiad}.}

\rvb{In the standard power system engineering context, transient stability typically requires a post-fault system to converge asymptotically to a stable equilibrium following large disturbances. Lagrange stability, on the other hand, is a mathematical property ensuring that the post-fault state trajectory remains bounded and is ultimately confined to a residual neighborhood around the SEP. It serves as a necessary but not sufficient condition for asymptotic stability, as it theoretically permits non-convergent-yet-bounded oscillations, such as limit cycles or chaotic behaviors.} 

\rvb{Consequently, ultimate boundedness alone therefore cannot certify convergence to the SEP and must be combined with a contraction argument on the residual radius itself. In the proposed framework, the residual radius $r$ is not a fixed quantity but contracts to zero as the trajectory approaches the SEP, as shown in the analysis of $r_{\max}$ below. The ultimate-boundedness conclusion therefore strengthens into asymptotic convergence to the SEP, so the energy-function argument delivers transient stability in the standard engineering sense, i.e., asymptotic convergence to the post-fault SEP following large disturbances. Additionally, in inverter-dominated systems, sustained oscillations can additionally arise from inverters repeatedly hitting their current limiters. Such scenario is not covered by the energy-function argument developed in this paper, which points to a meaningful direction for future work. }

The bounded region $\Omega$ is defined through the condition $\mathbf{Q_s}<\mathbf{0}$ as:
\begin{equation}
\mathbf{\Lambda}_\mathbf{L} \mathbf{D}_\mathbf{L} + (\mathbf{\Lambda}_\mathbf{L} \mathbf{D}_\mathbf{L})^\top - \frac{ (\mathbf{\Lambda}_\mathbf{L} \mathbf{D}_\mathbf{LM} ) \mathbf{D}_\mathbf{M}^\mathbf{-1}  ( \mathbf{\Lambda}_\mathbf{L} \mathbf{D}_\mathbf{LM} )^\top }{2} > \mathbf{0}
\label{DampingCond}
\end{equation}
where within $\Omega$, $\bm{\delta}_L$ and $\bm{\delta}_M$ satisfy the condition (\ref{DampingCond}) and $\bar{\bm{\omega}}_L \in \mathbb{R}^n$, $\bar{\bm{\omega}}_M \in \mathbb{R}^m$. Condition (\ref{DampingCond}) jointly leveraging the damping contribution from GFM inverters ($ \mathbf{D}_\mathbf{M}$) and explicitly accounts for the combined damping interactions ($\mathbf{D}_\mathbf{L}$ and $\mathbf{D}_\mathbf{LM}$). 

\rvb{Mathematically, (35) dictates that monotonic energy decay ($\dot{V}<0$) is mathematically guaranteed only on the intersection defined by $\Omega_\text{dec} := \Omega \cap \{\|\bar{\bm\omega}_L\|_2>r\}$. However, by analyzing the quadratic form, it can be demonstrated that under practical GFM inverter damping settings, this radius $r$ is negligibly small. Specifically, by applying the triangle inequality $\big\|\bar{\bm\omega}_L+\tfrac{1}{2}\mathbf{Q}_s^{-1}\mathbf{\Gamma}\big\|_2 \ge \|\bar{\bm\omega}_L\|_2-\|\mathbf\Gamma\|_2/(2\lambda_{\min}(-\mathbf{Q}_s))$ along with the matrix property $-\mathbf{Q}_s \succeq \lambda_{\min}(-\mathbf{Q}_s)\mathbf{I}$, we can confirm that $\dot{V}<0$ holds whenever $\|\bar{\bm\omega}_L\|_2>r$. The explicit closed-form upper bound for $r$ is derived as:
\begin{equation}
\begin{aligned}
    r \le r_{\max}& =\frac{\|\mathbf\Gamma\|_2}{\lambda_{\min}(-\mathbf{Q}_s)}
\\
&=\frac{2\,\|\mathbf\Gamma\|_2}{\lambda_{\min}\!\left[
\begin{aligned}
    &\mathbf{\Lambda}_\mathbf{L}\mathbf{D}_\mathbf{L}+(\mathbf{\Lambda}_\mathbf{L}\mathbf{D}_\mathbf{L})^{\!\top}-\\
    &\tfrac{1}{2}(\mathbf{\Lambda}_\mathbf{L}\mathbf{D}_\mathbf{LM})\mathbf{D}_\mathbf{M}^{-1}(\mathbf{\Lambda}_\mathbf{L}\mathbf{D}_\mathbf{LM})^{\!\top}
\end{aligned}\right]},
\end{aligned}
\label{r_bound}
\end{equation}
where, according to (\ref{eq:Gamma})--(\ref{eq:Gamma2}) and (\ref{eq:AB}), the residual vector $\mathbf\Gamma$ satisfies: 
\begin{equation}
\|\mathbf\Gamma\|_2 \le \sqrt{\,\sum_i\!\bigg[\,\sum_{j\neq i}2\,\big|(\mathbf{Y}_L^{-1})_{ij}\big|\,\|\mathbf{I}_{Lj}\|\,\Big|\sin\tfrac{\delta_{Lj}-\delta_{Lj}^{s}}{2}\Big|\,\bigg]^{\!2}\,}.
\label{Gamma_bound}
\end{equation}}

\rvb{As indicated by (\ref{r_bound}), the bounding radius $r_{\max}$ is tightly governed by the ratio of the network coupling (numerator) to the net damping margin (denominator). The numerator $\|\mathbf\Gamma\|_2$ is fundamentally driven by the GFL–GFL transfer impedance terms $|(\mathbf{Y}_L^{-1})_{ij}|$ ($i\ne j$). Because the network admittance matrix $\mathbf{Y}_L$ typically exhibits strong diagonal dominance, these off-diagonal transfer impedances remain inherently small. }

\rvb{Conversely, the denominator $\lambda_{\min}(-\mathbf{Q}_s)$ physically encapsulates the net equivalent damping margin of the mixed system. As evident in (\ref{DampingCond}), this margin is dictated by the primary damping interactions ($\mathbf{D}_\mathbf{L}$) minus a subtractive penalty term proportional to $\mathbf{D}_\mathbf{M}^{-1}$. Under typical engineering settings where GFM inverters provide sufficient virtual damping, the norm $\|\mathbf{D}_\mathbf{M}^{-1}\|$ is moderate. Consequently, the destabilizing penalty is well-controlled, ensuring that the net damping margin remains robustly positive, where a detailed analytical decomposition of these specific damping effects is provided subsequently in Section III-C. Because the numerator is suppressed by the network structure and the denominator is safeguarded by adequate GFM damping, the magnitude of $r_{\max}$ is mathematically minimized.} 

\rvb{Furthermore, as the inverter trajectories enter the sign-indefinite region where $\|\bar{\bm\omega}_L\|_2 \le r$, the structure of \eqref{Gamma_bound} ensures that each summand remains proportional to $\big|\sin\tfrac{\delta_{Lj}-\delta_{Lj}^{s}}{2}\big|$, it follows directly that
\begin{equation*}
\|\bm\Gamma\|_2 \;\longrightarrow\; 0 \quad \text{as} \quad \delta_{Lj}\to\delta_{Lj}^{s}, \;\;\forall j,
\end{equation*}
which in turn forces the bounding radius to vanish $r_{\max} \to 0$. Because this indefinite region dynamically collapses as the system stabilizes, the effective decay region $\Omega_\text{dec}$ practically coincides with the estimated region $\Omega$ (i.e., $\Omega_\text{dec} \approx \Omega$), validating the applicability of (\ref{DampingCond}) for stability assessment.}

\rvb{It should be noted that if the GFM virtual damping is reduced significantly below practical operating limits, the norm $\|\mathbf{D}_M^{-1}\|$ expands, driving a corresponding growth in the radius $r_{\max}$. In such a severely under-damped regime, which is rarely encountered in practical GFM deployments, $\Omega_{\mathrm{dec}}$ shrinks to a strict subset of $\Omega$, necessitating detailed time-domain simulations for an accurate transient stability assessment.}




Based on the energy function theory, \rvb{the critical energy is set as the minimum of two candidate quantities: the energy at the UEP inside $\Omega$ closest to the SEP, and the lowest energy attained on the boundary $\partial \Omega$ of the analysis region defined by~(36):}
\begin{equation}
\rvb{V_{cr} = \min\left\{V_{\text{UEP}}(\bm{\delta}_L^u, \bm{\delta}_M^u, \mathbf{0}, \mathbf{0}),\; \min_{x \in \partial \Omega} V(x)\right\},}
\label{Vcr}
\end{equation}
where $(\bm{\delta}_L^u, \bm{\delta}_M^u, \mathbf{0}, \mathbf{0})$ denotes the UEP closest to the SEP \rvb{and $\partial \Omega$ denotes the boundary of $\Omega$}. The boundary of the estimated stability region $\Omega_{es}$ is \rvb{then} set as the constant energy surface \rvb{defined by $V_{cr}$}:
\begin{equation}
\Omega_{es}=\{({\bm{\delta}_{L}},{\bm{\delta}_{M}},{{\bar{\bm{\omega}}}_{L}},{{\bar{\bm{\omega}}}_{M}}) | V \leq V_{cr}\}.
\label{Omega_es}
\end{equation}
\rvb{With this definition, the containment $\Omega_{es} \subseteq \Omega$ holds by construction.}

\rvb{The composite definition in~(\ref{Vcr}) extends the classical closest-UEP method to the local-energy-function setting employed in this paper. In the classical energy-function based framework, $V_{cr}$ is defined via the closest UEP where the energy function is globally decreasing. Our formulation, by contrast, constructs a local energy function on $\Omega$ and introduces an additional boundary $\partial \Omega$ defined by the damping condition~(36), which the classical framework does not address. As noted in~\cite[Section~6.5]{Chiang}, the utility of local analytical energy functions for lossy power systems remains an open question. By taking the minimum of the two candidate boundaries in~(\ref{Vcr}), the proposed framework explores this open direction and provides a natural extension of the closest-UEP framework to the local-energy-function setting.}

\rvb{Moreover, while taking the minimum ensures mathematical rigor, it is worth noting that under practical engineering conditions, $\min_{\mathbf{x} \in \partial \Omega} V(\mathbf{x})$ is inherently large. Specifically, the location of the boundary $\partial \Omega$ is fundamentally pushed outward by the GFM virtual damping. As shown in~(36), the matrix $\mathbf Q_s$ that defines $\Omega$ contains a destabilizing penalty term proportional to $\mathbf D_M^{-1}$. Larger GFM damping $\mathbf D_M$ therefore shrinks this penalty, leaving $\mathbf Q_s$ more strongly negative definite at the SEP. Since $\partial \Omega$ is reached only when $\mathbf Q_s$ loses negative definiteness, a stronger margin at the SEP requires larger angle deviations before this margin is exhausted, so $\partial \Omega$ sits at greater angular distance from the SEP. Consequently, $V_{\text{UEP}} < \min_{\mathbf{x} \in \partial \Omega} V(\mathbf{x})$ generally holds, and the closest UEP remains the active physical constraint in the criterion.}

\rvb{Notably, the closest UEP, rather than the controlling UEP, is adopted as the dynamical candidate in~(\ref{Vcr}).} Although it is well known that the controlling UEP may provide a less conservative stability boundary in some cases, systematic identification of the controlling UEP for high-order mixed GFM--GFL inverter systems is generally challenging. The closest UEP offers a clearly defined and computationally tractable reference for critical energy evaluation, which is consistent with the analytical objective of this paper. \rvb{The composite definition in~(\ref{Vcr}) is therefore deliberately conservative: by taking the minimum of two safety candidates and using the closest UEP for the dynamical candidate, the criterion errs on the safe side, in line with standard engineering practice for transient stability screening and protection coordination.}

\rvb{Accordingly, the system stability index regarding critical energy is defined as:}
\begin{equation}
\mu = (V_{cr}-V_{t_c})/V_{cr}
\end{equation}
where $V_{t_c}$ is the system energy calculated based on (\ref{Ly}) at the time $t_c$ when the disturbance is cleared. \rvb{It should be noted that the $\mu$-index does not claim classical asymptotic stability of the SEP from $\dot V<0$ alone. Strict decay $\dot V<0$ is established on the set $\Omega_{\mathrm{dec}} := \Omega \cap \{\|\bar{\bm\omega}_L\|_2 > r\}$, which certifies uniform ultimate boundedness of the post-fault trajectory with ultimate bound $r_{\max}$ in the sense of \cite[Thm.~4.18]{Khalil2002}. Asymptotic convergence to the SEP is recovered through a separate structural mechanism: the residual radius $r$ is not a fixed constant but a state-dependent quantity that vanishes as the trajectory approaches the SEP, since the residual vector $\bm\Gamma$ in (\ref{eq:Gamma}) encodes GFL angle deviations that are zero at the SEP by construction. The ultimate-boundedness ball therefore contracts onto the SEP along the post-fault trajectory, and the $\mu$-index serves as a sufficient certificate for the trajectory to enter this self-contracting region.}

At the SEP, $V_{t_c}=0$ and the corresponding stability index $\mu=1$. Following a large disturbance, if $V_{t_c}<V_{cr}$, i.e., $0<\mu<1$, the post-fault trajectory remains within the region $\Omega$ on which the energy-function argument applies, and \rvb{the system energy decays toward the residual level associated with the ultimate bound $r_{\max}$, which itself contracts to zero as the trajectory approaches the SEP.} Conversely, if $V_{t_c}$ exceeds $V_{cr}$, resulting in $\mu<0$, the trajectory cannot be confined to $\Omega$ and the system is deemed unstable after the disturbance.

The index $\mu$ thus serves as a quantitative measure of system stability following large disturbances. A large value of $\mu$ indicates a stronger capability for stability recovery, whereas a value of $\mu$ close to zero implies that the system is operating near the stability margin. 

\rvb{It is worth noting that the condition $V < V_{cr}$ provides a sufficient condition for transient stability rather than a necessary one. The local energy function is constructed to guarantee convergence to SEP only within a bounded region characterized by (40) with $r_{\max} \to 0$.} Outside this region, the time derivative of the energy function is not guaranteed to remain non-positive. As a result, the derived critical energy boundary is conservative and does not claim to exactly coincide with the true region of attraction of the system.

\rv{Despite this inherent conservativeness, in subsequent Section IV, extensive time-domain simulation results show that the critical clearing time (CCT) predicted by the proposed critical energy is very close to the actual CCT obtained from detailed nonlinear simulations. This indicates that, although conservative by design, the proposed critical energy is not overly pessimistic and can effectively capture the dominant transient synchronization mechanisms of the mixed GFM–GFL inverter system.}

\rv{Therefore, the proposed critical energy should be interpreted as a reliable and computationally efficient analytical indicator for post-fault transient stability assessment, rather than an exact characterization of the global stability region.}

\subsection{Effects of GFM Inverters' Damping on Transient Stability}
To further investigate the effect of GFM inverters' damping on system transient stability, rewrite $\mathbf{D}_\mathbf{L}$ as 
\begin{equation}
\mathbf{D}_\mathbf{L}=\tilde{\mathbf{D}}_\mathbf{L}+\tilde{\mathbf{D}}_{\mathbf{LM}}^{\bm{\Sigma}}
\end{equation}
where $\tilde{\mathbf{D}}_\mathbf{L},\ \tilde{\mathbf{D}}_{\mathbf{LM}}^{\bm{\Sigma}}\in {{\mathbb{R}}^{n\times n}}$:
\begin{equation}
 \tilde{\mathbf{D}}_\mathbf{L}(i,i)=\sum\limits_{j=1,j\ne i}^{n}{{{d}_{LLij}}}+{{d}_{Li}},\quad \tilde{\mathbf{D}}_\mathbf{L}(i,j)=-{{d}_{LLij}} 
\end{equation}
\begin{equation}
   \tilde{\mathbf{D}}_{\mathbf{LM}}^{\bm{\Sigma}}= \mathrm{diag}(\sum\limits_{j=1}^{m}{{{d}_{LM1j}}},\dots, \sum\limits_{j=1}^{m}{{{d}_{LMnj}}}).
\end{equation}

Then (\ref{DampingCond}) is further derived as:
\begin{equation} 
\begin{aligned}
&\mathbf{\Lambda}_\mathbf{L}\tilde{\mathbf{D}}_\mathbf{L}+\tilde{\mathbf{D}}_\mathbf{L}^\top \mathbf{\Lambda}_\mathbf{L}+\\
&\mathbf{\Lambda}_\mathbf{L}
\left[
  2\tilde{\mathbf{D}}_{\mathbf{LM}}^{\bm{\Sigma}}-
  \frac{ \mathbf{D}_\mathbf{LM} \mathbf{D}_\mathbf{M}^\mathbf{-1}  ( \mathbf{\Lambda}_\mathbf{L} \mathbf{D}_\mathbf{LM} )^\top }{2}  
 \right]> \mathbf{0}.
\end{aligned} \label{Condi2}
\end{equation}

Investigation of (\ref{Condi2}) reveals that since the damping of GFL inverters in mixed-GFM-GFL-inverter system ($\mathbf{D}_\mathbf{L}$, $\mathbf{D}_\mathbf{LM}$) is dynamically affected by the inter-inverter coupling, adding GFM inverters into the system does not necessarily enhance system transient stability. The effect is nontrivial and depends on whether the following term is positive definite:
\rv{\begin{equation}
2\mathbf{\Lambda}_\mathbf{L}\tilde{\mathbf{D}}_{\mathbf{LM}}^{\bm{\Sigma}}-\frac{ (\mathbf{\Lambda}_\mathbf{L}\mathbf{D}_\mathbf{LM}) \mathbf{D}_\mathbf{M}^\mathbf{-1}  ( \mathbf{\Lambda}_\mathbf{L} \mathbf{D}_\mathbf{LM} )^\top }{2} \label{GFMeff}
\end{equation}}
where $\mathbf{D}_\mathbf{LM}$ and $\mathbf{D}_{\mathbf{LM}}^{\bm{\Sigma}}$ are related to the electrical distance denoted by $\mathbf{K}_\mathbf{LM}$ and angle differences $\delta_{LMij}$ between GFL and GFM inverters. 

\rv{Under stable operating scenarios, $d_{LMij}$ can be rewritten based on (16b) as
\begin{equation*}
    {{d}_{LMij}}={{U}_{Mj}}\left| \mathbf{K}_{\mathbf{LM}(i,j)}\right| \cos (\delta_{Mij}+\phi_{ij}),
\end{equation*}
where $\sin \phi_{ij}=\mathbf{K}_{\mathbf{LM}(i,j)}^\mathbf{Im}/\left| \mathbf{K}_{\mathbf{LM}(i,j)}\right|$ and $\cos \phi_{ij}=-\mathbf{K}_{\mathbf{LM}(i,j)}^\mathbf{Re}/\left| \mathbf{K}_{\mathbf{LM}(i,j)}\right|$. The expression of $\mathbf{K}_{\mathbf{LM}}$ in (11a) can be further derived as:
\begin{equation*}
\begin{aligned}
\mathbf{K}_\mathbf{LM}&= \mathbf{K}_\mathbf{LM}^{\mathbf{Re}}+j\mathbf{K}_\mathbf{LM}^{\mathbf{Im}}=\mathbf{Y}_{\mathbf{L}}^{-1}\mathbf{Y}_{\mathbf{LM}}\\
&=(\mathbf{Y}_{\mathbf{L}}^\mathbf{{-1Re}} \cdot \mathbf{Y}_{\mathbf{LM}}^\mathbf{Re}-\mathbf{Y}_{\mathbf{L}}^\mathbf{{-1Im}} \cdot \mathbf{Y}_{\mathbf{LM}}^\mathbf{Im}) + \\
& \quad j (\mathbf{Y}_{\mathbf{L}}^\mathbf{{-1Im}} \cdot \mathbf{Y}_{\mathbf{LM}}^\mathbf{Re}+\mathbf{Y}_{\mathbf{L}}^\mathbf{{-1Re}} \cdot \mathbf{Y}_{\mathbf{LM}}^\mathbf{Im}).
\end{aligned}
\end{equation*}
Since transmission networks are typically reactance-dominated, it generally holds that $\left| \mathbf{Y}_{\mathbf{L}(i,j)}^\mathbf{{-1Re}} \right| \ll \left| \mathbf{Y}_{\mathbf{L}(i,j)}^\mathbf{{-1Im}} \right|$ and $\left|\mathbf{Y}_{\mathbf{LM}(i,j)}^\mathbf{Re} \right| \ll \left| \mathbf{Y}_{\mathbf{LM}(i,j)}^\mathbf{Im}\right|$. As a result, $\left\|\mathbf{Y}_{\mathbf{L}}^\mathbf{{-1Im}} \cdot \mathbf{Y}_{\mathbf{LM}}^\mathbf{Re}+\mathbf{Y}_{\mathbf{L}}^\mathbf{{-1Re}} \cdot \mathbf{Y}_{\mathbf{LM}}^\mathbf{Im} \right\| \ll \left\|\mathbf{Y}_{\mathbf{L}}^\mathbf{{-1Re}} \cdot \mathbf{Y}_{\mathbf{LM}}^\mathbf{Re}-\mathbf{Y}_{\mathbf{L}}^\mathbf{{-1Im}} \cdot \mathbf{Y}_{\mathbf{LM}}^\mathbf{Im}\right\|$. Consequently, it leads to a small $\phi_{ij}$ as $\left|\mathbf{K}_{\mathbf{LM}(i,j)}^\mathbf{Im}\right| \ll \left|\mathbf{K}_{\mathbf{LM}(i,j)}^\mathbf{Re}\right|$. Moreover, under normal operating conditions, $\delta_{LMij}$ is assumed to remain within the range $(-\pi/2,\pi/2)$, which is consistent with the conventional stable operating region \cite{Kundur1994Ch6}. Consequently, $\cos (\delta_{Mij}+\phi_{ij})$ remains positive. Therefore, each term $d_{LMij}$ is positive under normal operation, and the sum $\sum_{j=1}^{m}{{{d}_{LMij}}}$ is positive for all $i$. Since $\tilde{\mathbf{D}}_{\mathbf{LM}}^{\bm{\Sigma}}$ is a diagonal matrix
with strictly positive diagonal entries, it follows directly that $\tilde{\mathbf{D}}_{\mathbf{LM}}^{\bm{\Sigma}}$ is positive definite.}

\rv{As for the term $(\mathbf{\Lambda}_\mathbf{L}\mathbf{D}_\mathbf{LM}) \mathbf{D}_\mathbf{M}^\mathbf{-1}  ( \mathbf{\Lambda}_\mathbf{L} \mathbf{D}_\mathbf{LM} )^\top $, it is closely related to the magnitude of the coupling matrix $\mathbf{D}_\mathbf{LM}$, where $\mathbf{D}_{\mathbf{LM}}(i,j) = d_{LMij}$. The coefficient $ \mathbf{K}_{\mathbf{LM}(i,j)}$ in $d_{LMij}$ can be expressed as $\mathbf{K}_{\mathbf{LM}(i,j)}=(\mathbf{Y}_{\mathbf{L}}^{-1}\mathbf{Y}_{\mathbf{LM}})_{ij}=\sum_{k}\mathbf{Y}_{\mathbf{L}(i,k)}^{-1}\cdot \mathbf{Y}_{\mathbf{LM}(k,j)}$. In typical transmission networks, each coupling admittance $\mathbf{Y}_{\mathbf{LM}(k,j)}$ is usually much smaller compared with the self-admittance of the GFL subnetwork. Furthermore, since the admittance matrix $\mathbf{Y}_{\mathbf{L}}$ is strictly diagonally dominant, the elements of its inverse $\mathbf{Y}_{\mathbf{L}(i,k)}^{-1}$ are bounded. As a result, the summation of products of weak coupling terms and bounded coefficients leads to a small overall magnitude of $(\mathbf{Y}_{\mathbf{L}}^{-1}\mathbf{Y}_{\mathbf{LM}})_{ij}$, which implies $\left| \mathbf{K}_{\mathbf{LM}(i,j)} \right| \ll 1$ and thus $\left|d_{LMij}\right| \ll 1$. Since all elements of $\mathbf{D}_\mathbf{LM}$ are therefore small in magnitude, the quadratic term $(\mathbf{\Lambda}_\mathbf{L}\mathbf{D}_\mathbf{LM}) \mathbf{D}_\mathbf{M}^\mathbf{-1}  ( \mathbf{\Lambda}_\mathbf{L} \mathbf{D}_\mathbf{LM} )^\top $ also remains small.}

\rv{With $\tilde{\mathbf{D}}_{\mathbf{LM}}^{\bm{\Sigma}}$ being positive definite and the contribution of the quadratic term $(\mathbf{\Lambda}_\mathbf{L}\mathbf{D}_\mathbf{LM}) \mathbf{D}_\mathbf{M}^\mathbf{-1}  ( \mathbf{\Lambda}_\mathbf{L} \mathbf{D}_\mathbf{LM} )^\top $ being small, based on (\ref{condi}), it indicates that the presence of GFM inverters can accelerate the energy dissipation after large disturbances, thereby contributing positively to the stability of the mixed-GFM-GFL-inverter system.}

However, in the case where the GFM inverters are poorly damped, i.e., $\mathbf{D}_\mathbf{M}^\mathbf{-1}$ being relatively large, (\ref{GFMeff}) can become indefinite or even negative definite. Such a scenario undermines the fulfillment of condition (\ref{Condi2}) and may even make (\ref{Condi2}) negative definite. This suggests that through the inter-inverter coupling, poorly damped GFM inverters may decrease the energy dispassion rate and could even contribute to an increase in system energy. Such situation is particularly concerning because it implies that the addition of GFM inverters, under inappropriate damping settings, can degrade rather than enhance system stability, which compromises the ability of the GFM inverters to support network stability during transients. Therefore, appropriate tuning of GFM damping parameters is essential to ensure that their integration contributes constructively to the stability of the mixed GFM-GFL-inverter system.

\rv{\subsection{Effects of PLL Parameters on Transient Stability}}
\rv{As indicated in (\ref{Lambda_L}), the effects of PLL parameters are reflected in the matrix $\mathbf{\Lambda}_\mathbf{L}$. A preliminary inspection of (\ref{DampingCond}) shows that
\begin{equation}
    \mathbf{\Lambda}_\mathbf{L} \mathbf{D}_\mathbf{L} + (\mathbf{\Lambda}_\mathbf{L} \mathbf{D}_\mathbf{L})^\top \propto \mathbf{\Lambda}_\mathbf{L},
\end{equation}
while 
\begin{equation}
    \frac{ (\mathbf{\Lambda}_\mathbf{L} \mathbf{D}_\mathbf{LM} ) \mathbf{D}_\mathbf{M}^\mathbf{-1}  ( \mathbf{\Lambda}_\mathbf{L} \mathbf{D}_\mathbf{LM} )^\top }{2}\propto \mathbf{\Lambda}_\mathbf{L}^2.
\end{equation}
Since the second term scales quadratically with $\mathbf{\Lambda}_\mathbf{L}$, a overly large $\mathbf{\Lambda}_\mathbf{L}$ would make the inequality in (\ref{DampingCond}) harder to satisfy, thereby undermining the fulfillment of the sufficient damping condition.}

\rv{To conduct a more explicit investigation of the PLL parameters' effect, consider the simplified case where all GFL inverters share the same ratio $\lambda=K_{\text{P}i}^\text{PLL} /K_{\text{I}i}^\text{PLL}$, such that $\mathbf{\Lambda}_\mathbf{L}=\lambda \mathbf{I}$. Substituting this into (\ref{DampingCond}) yields
\begin{equation}
    \lambda \mathbf{D_L} + (\lambda \mathbf{D_L})^\top - \frac{1}{2} (\lambda \mathbf{D_{LM}}) \mathbf{D_M}^{-1} (\lambda \mathbf{D_{LM}})^\top > \mathbf{0}
\end{equation}
which can be rearranged as
\begin{equation}
\begin{aligned}
& \mathbf{D_L} + \mathbf{D_L^\top} - \frac{\lambda}{2}  \mathbf{D_{LM}} \mathbf{D_M}^{-1} \mathbf{D_{LM}^\top} > \mathbf{0}\\
 \Rightarrow & \lambda < \frac{2}{\lambda_{\max}((\mathbf{D_L} + \mathbf{D_L^\top})^{-1}\mathbf{D_{LM}} \mathbf{D_M}^{-1} \mathbf{D_{LM}^\top})}
\end{aligned}
\label{pll_pi}
\end{equation}
where $\lambda_{\max}$ denotes the maximum eigenvalue of the matrix.}

\rv{(\ref{pll_pi}) reveals that the ratio $K_{\text{P}i}^\text{PLL} /K_{\text{I}i}^\text{PLL}$ is subject to an upper bound determined by the system's dynamic damping, to ensure energy dissipation. Consequently, overly small $\mathbf{K}_{\mathbf{I}}^{\mathbf{PLL}}$ or overly large $\mathbf{K}_{\mathbf{P}}^{\mathbf{PLL}}$ would compromise the transient stability of the mixed GFM-GFL-inverter system. }

\section{Numerical Results}
\begin{figure}[!t]
\centering
\vspace{-0.1cm}
\includegraphics[width=3.5in]{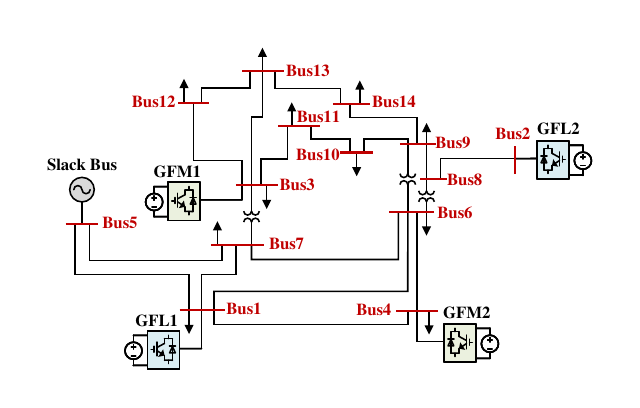}
\vspace{-0.5cm}
\caption{Standard IEEE 14-Bus test system incorporated with mixed GFM and GFL inverters.}
\vspace{-0.3cm}
\label{14bus}
\end{figure}

The test system with mixed GFM and GFL inverters is shown in Fig. \ref{14bus}. It has the same layout and line impedances as the standard IEEE 14-Bus test transmission system and the load is modeled as constant impedance load. The buses are renumbered and the SGs at Bus 1 $\sim$ Bus 4 are replaced by GFL and GFM inverters. The SG at Bus 5 is considered as the slack bus. The control parameters for the GFM and GFL inverters are listed in Table I.
\rv{During the fault-on period, the system is simulated using the full original network model. Upon fault clearing, the instantaneous post-fault system state is extracted. Accordingly, the inverter state variables are obtained and used to compute the system transient energy based on (21)–(24), which is then compared with the corresponding critical energy threshold to assess transient synchronization stability.}
 
\begin{table*}[!t]
\vspace{-0.5cm}
  \centering
    \begin{threeparttable}
    \caption{{Parameters of Mixed-GFM-GFL-Inverter System}}
    \vspace{-0.2cm}
    \label{tab1}
\begin{tabular}{m{3cm}<{\centering} m{10cm}<{\centering} m{3cm}<{\centering}}
    \toprule 
    \midrule 
Symbol & Description & Value \\
\midrule 
$U_g$ & Voltage magnitude of the slack bus  & 1 p.u.\\
$\omega_g$ & Ideal fixed angular speed of $xy$ common rotating reference  & $120\pi$ rad/s\\
$(P_{L1}^{\mathrm{ref}},Q_{L1}^{\mathrm{ref}})$ & Active and reactive power reference of GFL1 inverter & (0.5, 0.42) p.u.\\
$(P_{L2}^{\mathrm{ref}},Q_{L2}^{\mathrm{ref}})$ & Active and reactive power reference of GFL2 inverter & (0.5, 0.174) p.u.\\
$(P_{M1}^{\mathrm{ref}},U_{M1}^{d})$ & Active power and $d$-axis voltage reference of GFM1 inverter & (0.5, 1.05) p.u.\\
$(P_{M2}^{\mathrm{ref}},U_{M2}^{d})$ & Active power and $d$-axis voltage reference of GFM2 inverter & (0.5, 1.05) p.u.\\
$(K_{\text{P}1}^\text{PLL},K_{\text{I}1}^\text{PLL})$ & PLL proportional and integral gains of GFL1 inverter  & (0.2, 20) p.u.\\
$(K_{\text{P}2}^\text{PLL},K_{\text{I}2}^\text{PLL})$ & PLL proportional and integral gains of GFL2 inverter  & (0.25, 25) p.u.\\
$(J_1,D_1)$ & VSG Virtual inertia and damping coefficient of GFM1 inverter  & (1, 3) p.u.\\
$(J_2,D_2)$ & VSG Virtual inertia and damping coefficient of GFM2 inverter  & (1.5, 3.5) p.u.\\
\midrule 
    \bottomrule 
\end{tabular}
    \end{threeparttable}  
 \vspace{0.1cm}
\end{table*}

\subsection{Validation of System Critical Energy Value}

\begin{table}[!t]
\vspace{-0.1cm}
\centering
\caption{System SEP and UEP for Case 1, 2, 3}
\vspace{-0.2cm}
\begin{tabular}{m{1.6cm}<{\centering}  | m{3cm}<{\centering} m{3cm}<{\centering}}
\toprule
\midrule
($P_{Mi}^\mathrm{ref},P_{Li}^\mathrm{ref}$) p.u. &  SEP $(\bm{\delta}_L^s,\bm{\delta}_M^s,\bar{\bm{\omega}}_L^s,\bar{\bm{\omega}}_M^s)$ & UEP $(\bm{\delta}_L^u,\bm{\delta}_M^u, \bar{\bm{\omega}}_L^u, \bar{\bm{\omega}}_M^u)$\\
\midrule
$(0.5, 0.5)$ &{\shortstack{(-0.0593, -0.0271,\\-0.1401, -0.1659,\\0, 0, 0, 0)}} & 
{\shortstack{(-0.3388, -0.5111,\\-0.8277, -2.6713,\\0, 0, 0, 0)}}\\
\midrule
$(1.0, 0.5)$&{\shortstack{(-0.0153, 0.0703,\\0.0116, -0.0533,\\0, 0, 0, 0)}} & 
{\shortstack{(-0.2673, -0.1372,\\-0.4191, -2.6754,\\0, 0, 0, 0)}}\\
\midrule
$(0.5, 0.2)$ &{\shortstack{(-0.0870, -0.1517,\\-0.1894, -0.2033,\\0, 0, 0, 0)}} & 
{\shortstack{(-0.4039, -0.9801,\\-1.0559, -2.7227,\\0, 0, 0, 0)}}\\
\midrule
\bottomrule
\end{tabular} \label{SEPUEP}
\vspace{0.1cm}
\end{table}

\begin{table}[!t]
\vspace{-0.1cm}
\centering
\caption{System Energy Function under Different Fault Clearing Times}
\vspace{-0.2cm}
\begin{tabular}{m{1.6cm}<{\centering}  | m{1.1cm}<{\centering} m{1.1cm}<{\centering} m{1.1cm}<{\centering} m{1.2cm}<{\centering} }
\toprule
\midrule
($P_{Mi}^\mathrm{ref},P_{Li}^\mathrm{ref}$) p.u. &  $t_c$ (s) & $V_{t_c}$ (p.u.) & $V_{cr}$ (p.u.)& Stability \\
\midrule
\multirow{3}{*}{$(0.5, 0.5)$} & 0.074 & 7.709 & \multirow{3}{*}{7.877} & Stable \\
& 0.075 & 7.945 &   & Critical\\
& 0.076 & \textcolor{red}{8.182} &   & \textcolor{red}{Unstable}  \\
\midrule
\multirow{3}{*}{$(1.0, 0.5)$} & 0.102  &  9.300 & \multirow{3}{*}{9.535} & Stable \\
&  0.107  & 10.299  &   & Critical  \\
&  0.108 & \textcolor{red}{10.500} &   & \textcolor{red}{Unstable}  \\
\midrule
\multirow{3}{*}{$(0.5, 0.2)$} & 0.068  &  7.344 & \multirow{3}{*}{7.596} & Stable \\
& 0.069 &  7.592 &  & Critical  \\
& 0.070  &  \textcolor{red}{7.841} &  & \textcolor{red}{Unstable}  \\
\midrule
\bottomrule
\end{tabular}\label{Case123CCT}
\vspace{0.1cm}
\end{table}

To verify the effectiveness of the derived system critical energy value, three cases with different inverter active power set points are considered: Case 1: $P_{Mi}^{\mathrm{ref}}=0.5$ p.u., $P_{Li}^{\mathrm{ref}}=0.5$ p.u.; Case 2: $P_{Mi}^{\mathrm{ref}}=1$ p.u., $P_{Li}^{\mathrm{ref}}=0.5$ p.u.; and Case 3: $P_{Mi}^{\mathrm{ref}}=0.5$ p.u., $P_{Li}^{\mathrm{ref}}=0.2$ p.u.. The SEP and UEP for Case 1, Case 2 and Case 3 are summarized in Table \ref{SEPUEP}. In all three cases, the fault scenario is designed such that the voltage magnitude of the slack bus $U_g$ suddenly drops to 0 p.u., followed by its restoration to 1 p.u. after a fault clearing time of $t_c$ seconds. For each case, system energy at different fault clearing times, including the critical fault clearing time, is calculated and compared against the critical energy value $V_{cr}$ as shown in Table \ref{Case123CCT}.

\rv{Based on time-domain simulations, the CCT is identified as the maximum fault duration the system can sustain without losing synchronism, determined with a precision of 0.001 s.
It can be seen from Table \ref{Case123CCT} that, for Case 1 and Case 2, the system energy function already exceeds $V_{cr}$ at the CCT. And for Case 3, the system remains stable at a clearing time of 0.069 s but loses stability at 0.070 s. This indicates that the theoretical exact CCT lies between these two points. To maintain a conservative and practical stability assessment, we reported the last stable point (0.069 s) as the CCT. At this specific instant, since the fault duration is slightly shorter than the theoretical critical limit, the accumulated transient energy $V_{tc}$ is slightly lower than, yet already very close to, the critical threshold $V_{cr}$. Collectively, these three cases substantiate the effectiveness of the proposed critical energy threshold as a reliable stability indicator.} When the fault clearing time slightly exceeds the CCT, resulting in system instability, the corresponding system energy remains greater than $V_{cr}$, verifying the effectiveness of the proposed method. Conservatively, when the system energy is very close to, yet remains below, the critical energy $V_{cr}$, the corresponding fault clearing time is considered as the estimated CCT. \rv{From Table \ref{Case123CCT}, it can be observed that the estimated CCTs for Case 1, Case 2 and Case 3 are 0.074 s, 0.102 s and 0.068 s, respectively. These estimates are close to the actual CCT for the studied cases (0.075 s, 0.107 s and 0.069 s). The small deviations indicate that the proposed method exhibits a small conservativeness and provides sufficiently accurate stability assessments.}

\subsection{System Dynamics with Varied $D_i$ under Different Fault Types}

\begin{table*}[!t]
\vspace{-0.5cm}
\centering
\begin{threeparttable}
\caption{System Responses with Varied $D_i$ under Large Disturbances ($V_{cr}=7.877\  \mathrm{p.u.}$ )}
\vspace{-0.2cm}
\begin{tabular}{m{1.4cm}<{\centering} | m{1.2cm}<{\centering} | m{1cm}<{\centering} m{1cm}<{\centering} m{1cm}<{\centering} m{1.2cm}<{\centering} | m{1.2cm}<{\centering} | m{1cm}<{\centering} m{1cm}<{\centering} m{1cm}<{\centering} m{1.2cm}<{\centering}}
\toprule
\midrule
($D_1,D_2$) p.u. & Fault type & $t_c$ (s) & $V_{t_c}$ (p.u.)& $\mu$ & Stability & Fault type & $t_c$ (s) & $V_{t_c}$ (p.u.) & $\mu$ & Stability \\
\midrule
\multirow{4}{*}{$(0.3, 0.35)$} & \multirow{16}{*}{\shortstack{Voltage\\sag of\\slack bus}} & 0.050 & 3.309 & 0.580 & Stable & \multirow{16}{*}{\shortstack{Three\\-phase\\short-\\circuit\\at Bus 6}} & 0.050 & 3.522 & 0.553 & Stable \\
& & 0.070 & 7.585 & 0.037 & Stable & & 0.070 & 6.497 & 0.175 & Stable \\
& & 0.071 & 7.844 & 0.004 & Critical & & 0.082 & 8.400 & \textcolor{red}{-0.066} & Critical \\
& & 0.072 & 8.106 & \textcolor{red}{-0.029} & \textcolor{red}{Unstable} & & 0.083 & 8.558 & \textcolor{red}{-0.086} & \textcolor{red}{Unstable} \\
\cline{1-1} \cline{3-6} \cline{8-11}
\multirow{4}{*}{$(1.0, 1.5)$} & & 0.050 & 3.208 & 0.593 & Stable & & 0.050 & 3.438 & 0.564 & Stable \\
& & 0.070 & 7.302 & 0.073 & Stable & & 0.070 & 6.292 & 0.212 & Stable \\
& & 0.072 & 7.800 & 0.010 & Critical & & 0.084 & 8.404 & \textcolor{red}{-0.067} & Critical \\
& & 0.073 & 8.053 & \textcolor{red}{-0.022} & \textcolor{red}{Unstable} & & 0.085 & 8.554 & \textcolor{red}{-0.086} & \textcolor{red}{Unstable} \\
\cline{1-1} \cline{3-6} \cline{8-11}
\multirow{4}{*}{$(3.0, 3.5)$} & & 0.050 & 3.029 & 0.615 & Stable & & 0.050 & 3.298 & 0.581 & Stable \\
& & 0.070 & 6.789 & 0.138 & Stable & & 0.070 & 5.955 & 0.244 & Stable \\
& & 0.075 & 7.945 & \textcolor{red}{-0.009} & Critical & & 0.088 & 8.440 & \textcolor{red}{-0.071} & Critical \\
& & 0.076 & 8.182 & \textcolor{red}{-0.039} & \textcolor{red}{Unstable} & & 0.089 & 8.576 & \textcolor{red}{-0.089} &\textcolor{red}{Unstable} \\
\cline{1-1} \cline{3-6} \cline{8-11}
\multirow{4}{*}{$(10, 10.5)$} & & 0.050 & 2.496 & 0.683 & Stable & & 0.050 & 2.874 & 0.635 & Stable \\
& & 0.070 & 5.331 & 0.138 & Stable & & 0.070 & 5.955 & 0.244 & Stable \\
& & 0.087 & 8.353 & \textcolor{red}{-0.060} & Critical & & 0.104 & 8.401 & \textcolor{red}{-0.067} & Critical \\
& & 0.088 & 8.535 & \textcolor{red}{-0.084} & \textcolor{red}{Unstable} & & 0.105 & 8.496 & \textcolor{red}{-0.079} & \textcolor{red}{Unstable} \\
\midrule
\bottomrule
\end{tabular}
\end{threeparttable}  
\vspace{0.1cm}
\label{AllResponse}
\end{table*}


\begin{figure}[!t]
\centering
\vspace{-0.1cm}
\includegraphics[width=3in]{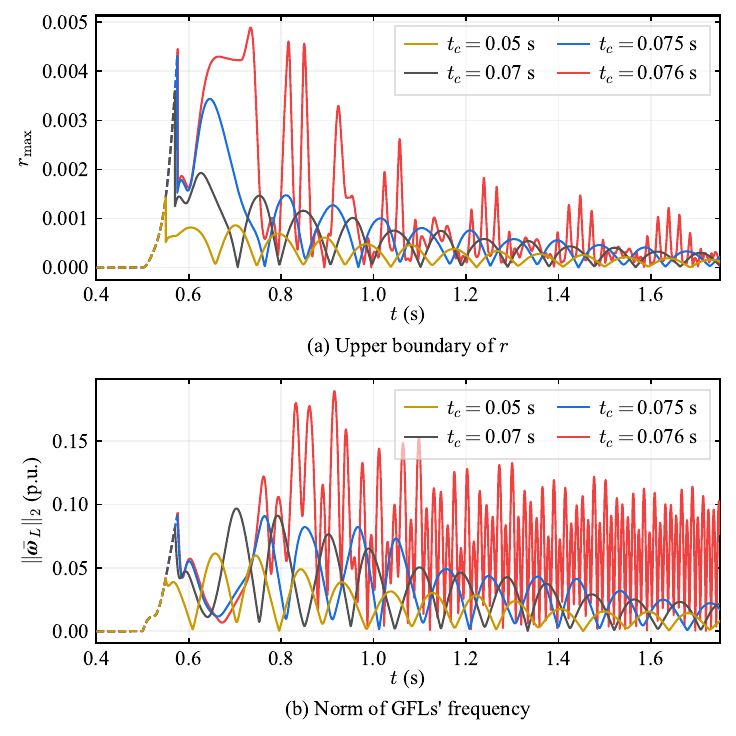}
\vspace{-0.5cm}
\caption{\rvb{Comparison of $r_{\max}$ and $\|\bar{\bm{\omega}}_L\|_2$ for system under $U_g$ sag disturbance with different fault duration times.}}
\vspace{-0.3cm}
\label{compare2}
\end{figure}

\begin{figure}[!t]
\centering
\vspace{-0.1cm}
\includegraphics[width=3.6in]{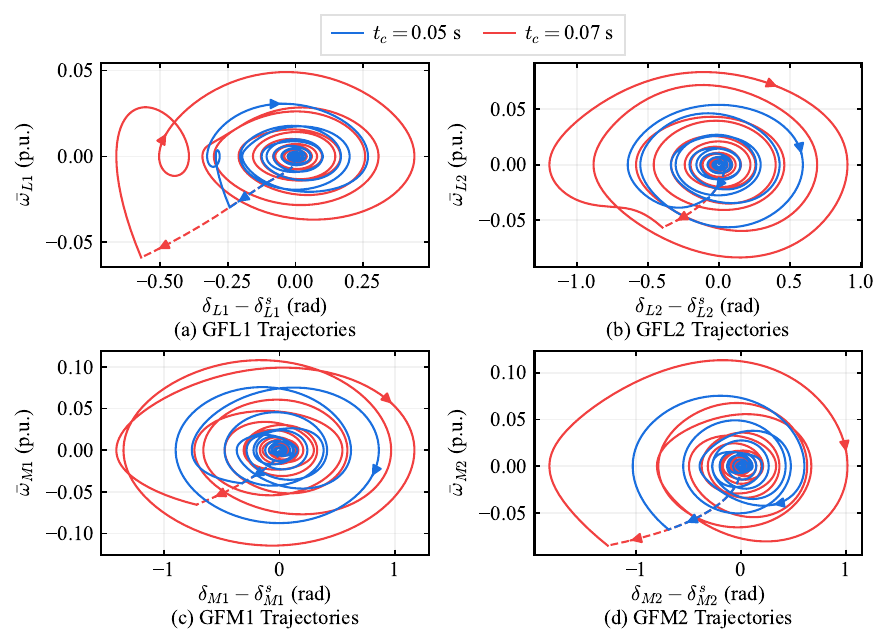}
\vspace{-0.5cm}
\caption{$U_g$ sag with $(D_1,D_2)=(3.0,3.5)$ p.u. under $t_c=0.050$ s and $t_c=0.070$ s.}
\vspace{-0.3cm}
\label{Case1M1_1}
\end{figure}
\begin{figure}[!t]
\centering
\vspace{-0.1cm}
\includegraphics[width=3.6in]{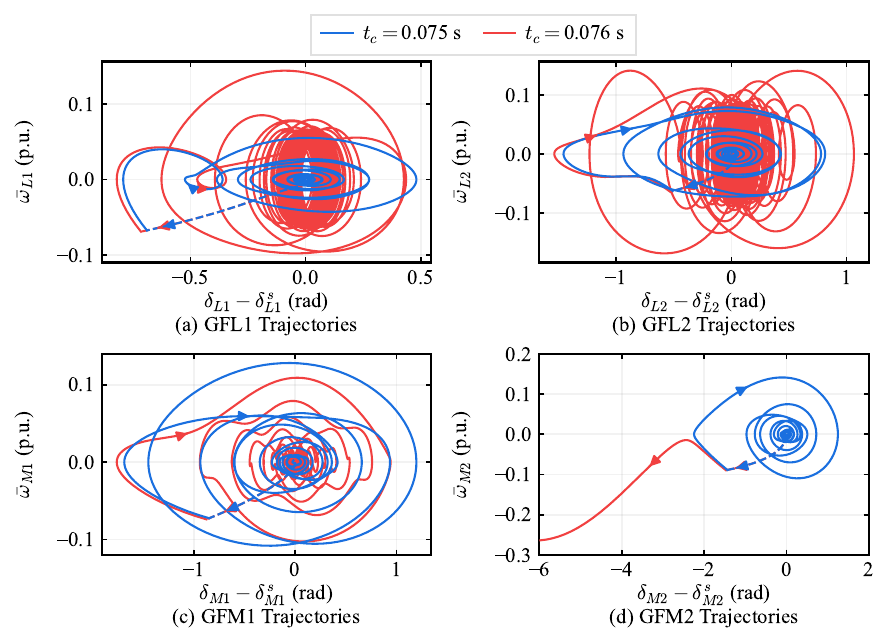}
\vspace{-0.5cm}
\caption{$U_g$ sag with $(D_1,D_2)=(3.0,3.5)$ p.u. under $t_c=0.075$ s and $t_c=0.076$ s.}
\vspace{-0.3cm}
\label{Case1M1_2}
\end{figure}

\begin{figure}[!t]
\centering
\vspace{-0.1cm}
\includegraphics[width=3in]{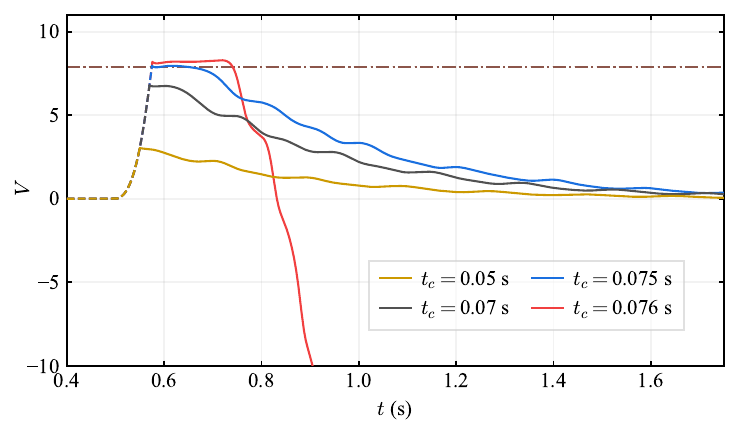}
\vspace{-0.5cm}
\caption{Energy function with $(D_1,D_2)=(3.0,3.5)$ p.u. $t_c=0.050$ s, $t_c=0.070$ s, $t_c=0.075$ s and $t_c=0.076$ s.}
\vspace{-0.3cm}
\label{VCase1M1}
\end{figure}

To further exam the impact of the GFM inverters' damping coefficients on system transient stability, additional simulations are conducted under Case 1 with varied $D_i$. Specifically, system dynamics are analyzed for four damping settings: (1) $(D_1, D_2)=(0.3, 0.35)$ p.u.; (2) $(D_1, D_2)=(1.0, 1.5)$ p.u.; (3) $(D_1, D_2)=(3.0, 3.5)$ p.u. and (4) $(D_1, D_2)=(10, 10.5)$ p.u.. Two types of disturbances are considered: the sudden voltage sag of $U_g$ and three-phase short-circuit at Bus 6. The corresponding system responses are summarized in Table IV and the dynamic behaviors for case $(D_1, D_2)=(3.0, 3.5)$ p.u. are illustrated in Fig. \ref{compare2} - Fig. \ref{VCase1M1} for detailed interpretation.

\rvb{Specifically, to validate that the indefinite region of $V$ is highly localized as discussed in Section III-B, Fig. \ref{compare2} compares the real-time value of $r_{\max}$ and the $\|\bar{\boldsymbol{\omega}}_L\|_2$ tracked along the post-fault trajectories for the detailed case under different fault duration times. As shown in Fig. \ref{compare2}, the region where strict energy decay is not guaranteed, occurring only when $\|\bar{\bm\omega}_L\ \le r_{\max}\|$, constitutes a minuscule fraction of the overall post-fault frequency excursions. Consequently, the indefinite inner region $\Omega \cap\{\|\bar{\bm\omega}_L\|\le r\}$ represents, in practice, an extremely tight neighborhood around the SEP that the trajectory only enters during the very final stage of system recovery.}

It is evident from Table IV that across all scenarios, an increase in fault clearing time leads to a corresponding increase in system energy at the clearing time and a decrease in the stability index $\mu$, indicating a reduced ability of the system to maintain stability. It can be further observed from Table IV that, in all unstable scenarios, $\mu<0$, thereby validating the effective of the proposed method in correctly identifying instability. In addition, for all cases involving three-phase short-circuit faults, as well as for the cases with damping settings $(D_1, D_2)=(3.0, 3.5)$ p.u. and $(D_1, D_2)=(10, 10.5)$ p.u. under voltage sag faults, the system energy at CCT also exceeds the critical energy threshold. This outcome highlights the conservative nature of the proposed method, ensuring a margin of safety in stability assessment.

By comparing the results at the same fault clearing times of 0.050 s and 0.070 s with varied $(D_1,D_2)$, it is observed that increasing $(D_1,D_2)$ from $(0.3, \ 0.35)$ p.u. to $(10, \ 10.5)$ p.u. leads to a measurable improvement in stability index. Specifically, for $t_c=0.050$ s, $\mu$ increases from 0.580 to 0.683 for the $U_g$ sag fault and from 0.553 to 0.635 for the short-circuit fault. Similarly, for $t_c=0.070$ s , $\mu$ increases from 0.037 to 0.138 and from 0.175 to 0.244 for the two respective fault types. These findings align with the analysis derived from (\ref{GFMeff}), confirming that enhanced damping in GFM inverters exerts a stabilizing influence on the system. 

Specifically, as shown in Fig. \ref{Case1M1_1} and Fig. \ref{Case1M1_2} for $(D_1, D_2)=(3.0,\  3.5)$ p.u., the inverter trajectories converge to the SEP under $t_c=0.050$ s and  $t_c=0.070$ s, indicating system stability. At the CCT $t_c=0.075$ s, the system can still converge to the SEP. While with a slight increase to $t_c=0.076$ s, the trajectory of GFM 2 inverter, shown by the red line in Fig. \ref{Case1M1_2} (d), diverges from the SEP and is the first to lose stability. Subsequently, due to inter-inverter coupling effects, GFL 1, GFL 2, and GFM 3 inverters' trajectories, as depicted in Fig. \ref{Case1M1_2} (a)–(c), initially exhibit a tendency to converge but continue oscillating around the SEP without achieving asymptotic stability.

Correspondingly, the system energy dynamics, as shown in Fig. \ref{VCase1M1}, further validate the stability behavior observed in the phase trajectories. The critical energy value $V_{cr}$ is marked by a brown dash-dotted line. As shown by the solid yellow and black lines, the system energy remains below $V_{cr}$ for $t_c=0.050$ s and  $t_c=0.070$ s and gradually decaying to zero over time. At the CCT (blue solid line, $t_c=0.075$ s), the system energy slightly exceeds $V_{cr}$ but ultimately converges to zero, demonstrating marginal stability. However, for $t_c=0.076$ s (red solid line), the system energy surpasses $V_{cr}$ and subsequently goes unbounded, confirming the loss of stability as the system states diverge. These energy dynamics are consistent with the phase trajectory observations in Fig. \ref{Case1M1_1} and Fig. \ref{Case1M1_2}. 

It is worth noting that the minor oscillations during energy decay arise from neglecting the linear term $\frac{1}{4}{{\mathbf{\Gamma} }^{T}}\mathbf{Q_s^{-1}}\mathbf{\Gamma}$ in (\ref{condi}). However, these oscillations do not effect the overall decreasing trend of the system energy, which in turn, substantiates the validity of omitting the linear term in the stability assessment.

\subsection{Adverse Effect of Insufficient GFM Inverter Damping on System Transient Stability}
\begin{figure}[!t]
\centering
\vspace{-0.1cm}
\includegraphics[width=3.6in]{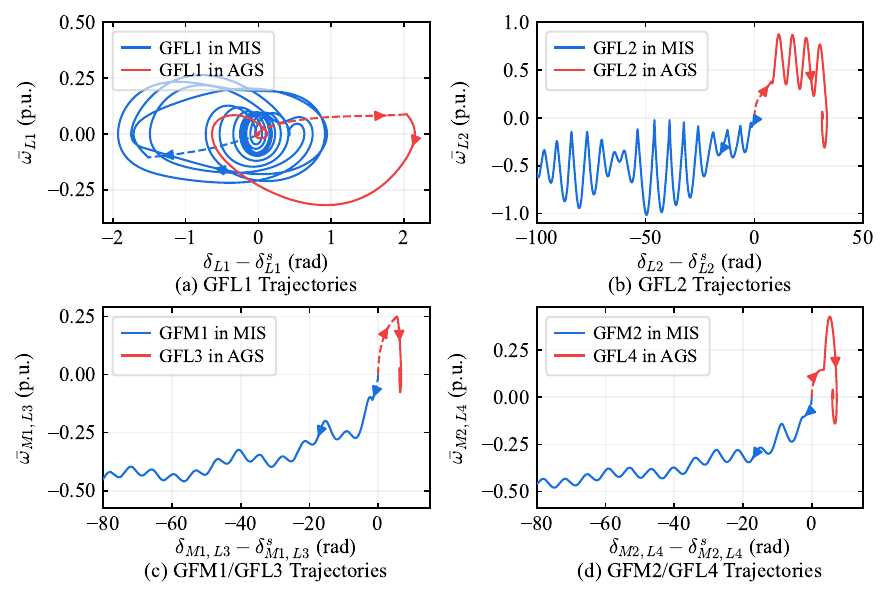}
\vspace{-0.5cm}
\caption{Mixed-GFM-GFL-inverter system and all-GFL-inverter system dynamics with $(D_1, D_2)=(3.0,3.5)$ p.u. under $U_g$ sag and $t_c=0.10$ s}
\vspace{-0.3cm}
\label{Compare1}
\end{figure}

\begin{figure}[!t]
\centering
\vspace{-0.1cm}
\includegraphics[width=3.6in]{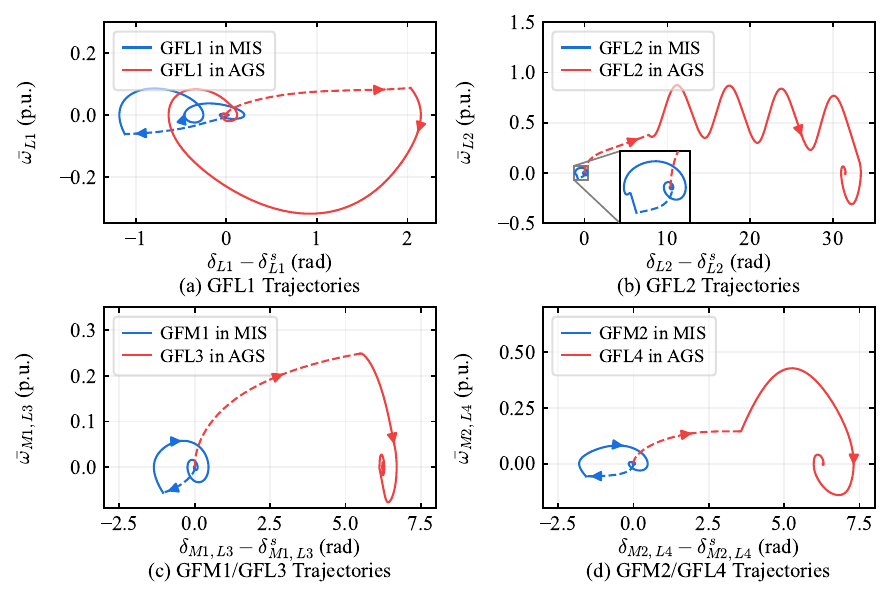}
\vspace{-0.5cm}
\caption{Mixed-GFM-GFL-inverter system and all-GFL-inverter system dynamics with $(D_1, D_2)=(20,25)$ p.u. under $U_g$ sag and $t_c=0.10$ s}
\vspace{-0.3cm}
\label{Compare2}
\end{figure}

Investigation of (\ref{GFMeff}) demonstrates that insufficient GFM inverter damping can adversely effect system transient stability, while properly increased $D_i$ can further enhance system stability. Accordingly, the GFM 1 and GFM 2 inverters located at Bus 3 and Bus 4 are replaced with GFL 3 and GFL 4 inverters, whose parameters are identical to those of GFL 1, to facilitate a comparison between the dynamics of a system composed entirely of GFL inverters and that of a mixed GFM–GFL-inverter system. The dynamic responses of the two configurations are compared in Fig. \ref{Compare1} and Fig. \ref{Compare2}, where MIS denotes the mixed-GFM–GFL-inverter system (blue lines) and AGS denotes the all-GFL-inverter system (red lines).

The comparative analysis in Fig. \ref{Compare1} reveals distinct stability behaviors between the mixed-GFM-GFL-inverter system and all-GFL-inverter system under the voltage sag fault ($U_g$ sag) with $t_c=0.10$ s. When $(D_1, D_2)=(3.0, 3.5)$ p.u. in the mixed-inverter configuration, GFM 1, GFM 2 and GFL 2 inverters lose stability, causing GFL 1 inverter's trajectory to persistently oscillate around the SEP. In contrast, the all-GFL-inverter system exhibits different post-fault dynamics: GFL 3 and GFL 4 inverters (which replace GFM 1 and GFM 2 inverters at the same buses) stabilize at the next SEP, while GFL 2 inverter converges to a fourth-period SEP. Notably, GFL 1 inverter in the all-GFL-inverter system successfully returns to its original SEP. These results validate that when GFM inverters are implemented with insufficient damping coefficients, they may fail to enhance stability compared to an all-GFL-inverter configuration, and can even introduce instability by preventing certain inverters from recovering to steady-state operation.

\begin{figure}[!t]
\centering
\vspace{-0.1cm}
\includegraphics[width=3.6in]{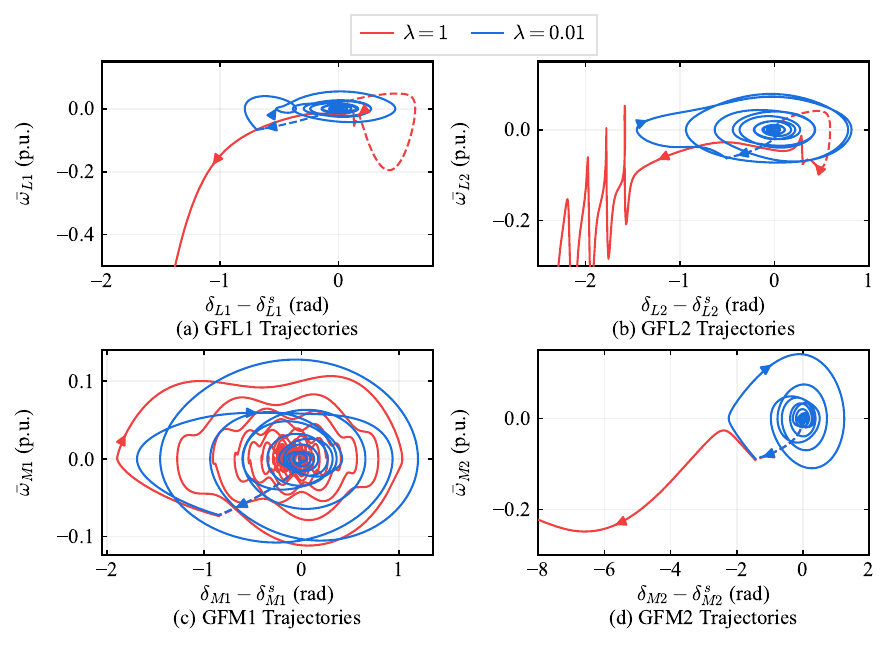}
\vspace{-0.5cm}
\rv{\caption{$U_g$ sag with $\lambda=1$ and $\lambda=0.01$ under $t_c=0.075$ s.}}
\vspace{-0.3cm}
\label{PLL_lambda1}
\end{figure}
\begin{figure}[!t]
\centering
\vspace{-0.1cm}
\includegraphics[width=3in]{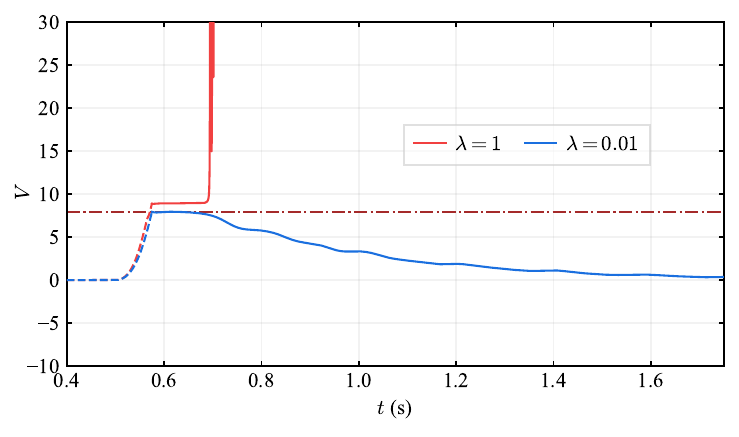}
\vspace{-0.5cm}
\rv{\caption{Energy function with $\lambda=1$ and $\lambda=0.01$ under $t_c=0.075$ s.}}
\vspace{-0.3cm}
\label{PLL_lambda2}
\end{figure}

As shown in Fig. \ref{Compare2}, when the GFM inverter damping coefficients in the mixed system are increased to $(D_1, D_2)=(20, 25)$ p.u., the system exhibits significantly improved transient stability under the same $U_g$ sag fault with $t_c=0.10$ s. Unlike the case with lower damping, all inverters in the mixed-GFM-GFL system now successfully converge to their original SEPs, demonstrating full post-fault recovery \rv{(MIS, blue lines)}. \rvb{In contrast, while the all-GFL-inverter system stabilizes, GFL 2, GFL 3 and GFL 4 inverters converge to different SEPs (AGS, red lines)}. This marked improvement in transient stability directly correlates with the enhanced damping capability of the GFM inverters, confirming that properly designed GFM converters with sufficient damping can significantly increase the system's stability to large disturbances. 

\rv{\subsection{Impacts of PLL Parameters on Transient Stability}}
\rv{Condition (\ref{pll_pi}) indicates that the overly large ratio $\lambda=K_{\text{P}i}^\text{PLL} /K_{\text{I}i}^\text{PLL}$ can deteriorate the transient stability of the mixed GFM-GFL-inverter system. To illustrate this effect, two representative scenarios are considered. Scenario 1 ($\lambda=1$) represents a high-ratio case with GFL parameters set to $K_{\text{P}1}^\text{PLL}=20, K_{\text{I}1}^\text{PLL}=20$ and $K_{\text{P}1}^\text{PLL}=25, K_{\text{I}1}^\text{PLL}=25$. Conversely, Scenario 2 ($\lambda=0.01$) represents a low-ratio case with $K_{\text{P}1}^\text{PLL}=0.2, K_{\text{I}1}^\text{PLL}=20$ and $K_{\text{P}1}^\text{PLL}=0.25, K_{\text{I}1}^\text{PLL}=25$. Throughout the simulations, the damping coefficients of the two GFM inverters are maintained at $(D_1, D_2)=(3.0, 3.5)$ p.u. }

\rv{Fig. \ref{PLL_lambda1} compares the system trajectory for the two scenarios under the same disturbance of $U_g$ sag with $t_c=0.075$ s. As indicated by the blue trajectories ($\lambda=0.01$), all inverters converge to SEP. However, when $\lambda$ is increased to 1, the system response changes substantially. As illustrated by the red trajectories in Fig. \ref{PLL_lambda1} (a) (b) and (d), GFL1, GFL2 and GFM2 inverters diverge from the SEP when the fault is cleared. Consequently, due to the coupling among the four inverters, GFM1 inverter's trajectory oscillates persistently around the SEP and fails to converge.}

\rv{Fig. \ref{PLL_lambda2} shows the corresponding system energy dynamics for the two PLL parameter scenarios. It can be observed that with $\lambda=0.01$, the system energy peaks near the critical value when the fault is cleared and subsequently decays to zero, indicating the marginal stability of the post-fault trajectories. In contrast, with $\lambda=1$, the system energy exceeds the critical threshold and diverges to infinity. This unbounded growth in energy is consistent with the diverging phase trajectories previously observed in Fig. \ref{PLL_lambda1}. These results validate that increasing $\lambda$ beyond a critical threshold compromises the system's ability to dissipate transient energy, leading to potential instability in the mixed GFM-GFL inverter system.}

\vspace{0.1cm}
\section{Conclusions}
This paper proposes a comprehensive transient stability analysis framework for mixed GFM–GFL-inverter systems. A detailed mathematical model is established, capturing the inter-inverter coupling mechanisms between GFM and GFL inverters during transient dynamics. By developing a local energy function for the system, the critical energy threshold is identified for fast stability assessment after large disturbances. The proposed stability index provides a quantitative approach to evaluate system stability. The analytical investigation further reveals the adverse effect of small GFM inverter damping coefficients on system transient stability. Extensive case studies on the IEEE 14-Bus test system further validate that the proposed method offers a reliable and computationally efficient tool for transient stability assessment in mixed-GFM-GFL-inverter system. For future work, the proposed framework will be extended to incorporate the impact of inner control loops, reactive power control dynamics, and fault ride-through strategies. \rvb{Specifically, to address the limitations of the current energy-function argument, future research will thoroughly investigate the repeated saturation dynamics introduced by inverter current limiters under severe faults and the limit cycle phenomena, to achieve a more comprehensive assessment of transient stability.} Moreover, the proposed method will be extended to black-box inverter systems with parameter identification to further enhance its practical applicability.

\bibliographystyle{IEEEtran}
\bibliography{paper}

\end{document}